# Plasma-Cascade micro-bunching Amplifier
# and Coherent electron Cooling of a Hadron Beams

V.N. Litvinenko[1,2], G. Wang[2,1], D. Kayran[2,1], Y. Jing[2,1], J. Ma[2] and I. Pinayev[2]
[1] Department of Physics and Astronomy, Stony Brook University, Stony Brook, NY
[2] Collider-Accelerator Department, Brookhaven National Laboratory, Upton, NY

*Abstract.* In this paper we describe an instability, which we called a Plasma-Cascade Amplifier (PCA), occurring in electron beams propagating along a straight trajectory. Such instability can strongly intensify longitudinal micro-bunching originating from the beam's shot noise, and even saturate it. Resulting random density and energy microstructures in the beam can become a serious problem for generating high quality electron beams.

On a positive side, the Plasma-Cascade micro-bunching amplifier can find multiple applications in light sources, for example, in generating high power broadband THz radiation. We discuss these topics in the last chapter the paper, while focusing on the application of this instability for cooling intense hadron beams.

Cooling high energy, high intensity hadron beams remains one of serious challenges in modern accelerator physics. A Coherent electron Cooling (CeC) is potentially a most promising technique to answer this challenge. All CeC schemes are based on enhancing electro-static interactions between electrons and hadrons, e.g. amplifying the microscopic imprints of hadrons in the electron beam density modulation. Three types of amplifiers had been proposed for CeC: a high-gain free-electron laser, a micro-bunching instability using magnetic chicanes and a hybrid scheme employing a broad-band laser. All of these schemes require bending of electron beam trajectory, which delay electrons with respect to the hadrons. Consequently, these schemes require separating and delaying the hadron beam, which both complex and very expensive.

We propose a CeC with PCA, which does not require any bending neither of electron nor hadron beams. It is centered on the exponential instability of longitudinal plasma oscillations driven by strong modulation of the electron beam density via controlling its transverse size.

In this paper we present detailed description of the PCA. We discuss a number CeC systems based on these scheme. We propose demonstrating this novel cooling mechanism using existing CeC system installed at Relativistic Heavy Ion Collider at Brookhaven National Laboratory.

While CeC is generally aimed toward cooling high energy hadron beams with energies, using the PCA makes it well suited for cooling beams hadron beam with modest energies from few to tens of GeV, for example in SIS-100 or SIS-300 at GSI.

## I. Introduction

In contrast to electron and positron beams, hadron beams in all present-day colliders do not have a strong natural damping mechanism to reduce their energy spreads and emittances. Cooling hadron beams transversely and longitudinally at the energy of collision may significantly increase the luminosity of high-energy hadron colliders and future electron-hadron colliders [1], such as RHIC [2], future US electron-ion collider (EIC) [3], and even the LHC/LHeC [4,5]. Furthermore, high-brightness high-energy hadron beams promise to become important drivers for hadron-based plasma accelerators [6].

Presently, two techniques are used for cooling hadron beams: electron cooling [7,8], and stochastic cooling [9,10]. Unfortunately, the efficiency of traditional electron cooling rapidly falls with an increase in the beam's energy and cooling proton beams at energy above 100 GeV is at least very challenging, if at all possible. The efficiency of traditional stochastic cooling, while independent of the particles' energy, rapidly falls with the increasing longitudinal density of particles [9]. Hence, while this technique was very successful with ion beams having particles densities of ~ $10^9$ per nanosecond [10], it is ineffective for proton and ion beams with linear density ~ $10^{11}$-$10^{12}$ protons per nanosecond, typical for modern hadron colliders.

Presently two potential candidates that might be up to the task: an optical stochastic cooling (OSC) [11], and a CeC (CeC) [12-16]. Both of these techniques belong to the family of stochastic coolers [9], but with the amplifier's bandwidth extending into optical region and beyond. The main advantage of CeC, originally suggested by Derbenev [17-19], is based in electrostatic interaction between ions and electrons. As the result, it is very flexible and can be used for variety of hadron beam energies. While the OSC technique is very interesting, it is unfortunately based upon a fixed-wavelength laser (amplifying undulator radiation from the hadron beam) - the feature making this approach inflexible. Hence, it is hardly useable, if at all, for hadron colliders operating at various energies. For example, cooling EIC hadron beam at 50 GeV and 250 GeV with the same OSC system would necessitate changing the amplifier's wavelength by a factor of 25, i.e., well beyond capabilities of current lasers. In contrast, the CeC technique is based on the fully adjustable optics-free mechanisms and the frequency amplifier naturally scales with the particles' energy.

Fig.1 shows schematics of CeC with various amplifiers. In the CeC both electron and hadron beams have identical relativistic factors $\gamma_o$ and velocities $v_o$:

$$\gamma_o = \frac{E_e}{m_e c^2} = \frac{E_h}{m_h c^2}; \; v_o = c\sqrt{1-\gamma_o^{-2}}, \qquad (1)$$

where $c$ is the speed of the light, $E_{e,h}$ and $m_{e,h}$ are the energies and masses of electrons and hadrons, correspondingly. The CeC works as follows: In the modulator, each hadron induces a density modulation in the electron beam by attracting surrounding electrons – the process called Debye screening in plasma physics. This modulation in electron density is boosted in the CeC amplifier, which has to operate in linear regime to preserve correlation between the amplified density modulation and the hadron initiating it. In other words, the resulting density perturbation in the electron beam would be a liner superposition of density spikes (or better say, wave-packets) induced by individual hadrons. In the kicker, the hadrons coherently interact with the self-induced electric field in the electron beam and receive energy kicks toward their central energy. The cooling mode is provided by placing a hadron with the nominal energy (1) on the crest of electron density modulation, which is induced by the hadron in the modulator and later amplified. Longitudinal electric field at this location is zero, and the hadron does not receive any energy kick. A hadron with higher energy arrives into the kicker in front of the high-density slice (shaded in darker blue in Figs. 1). It will be pulled backwards (decelerated) by the longitudinal eclectic field of the modulated electron beam. Vice versa, hadrons with lower than nominal energy would slip back and would be pulled forward (accelerated). This process reduces the hadron's energy spread, i.e., it cools the hadron beam longitudinally.

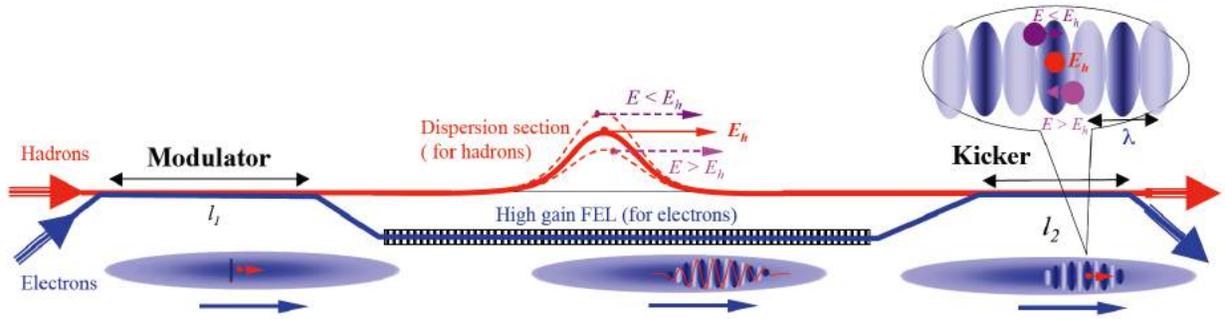

Figure 1.a. Schematic of the FEL-based Coherent Electron Cooler [12]. For clarity, the size of the FEL wavelength, λ, is exaggerated grossly.

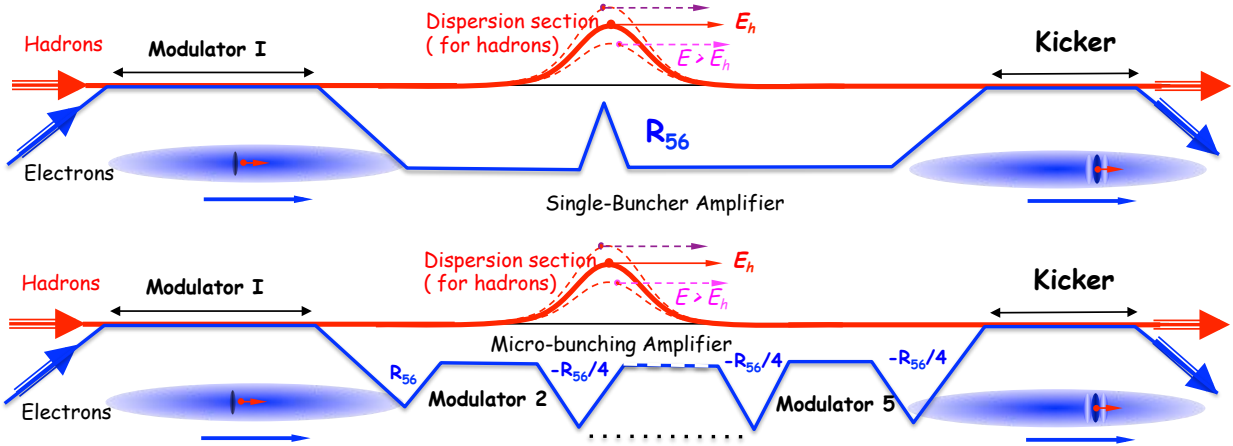

Figure 1.b. A CeC with a single stage (top) [14-16] and a multi-stage (bottom) high gain micro-bunching instability amplifier (MBIA) [14].

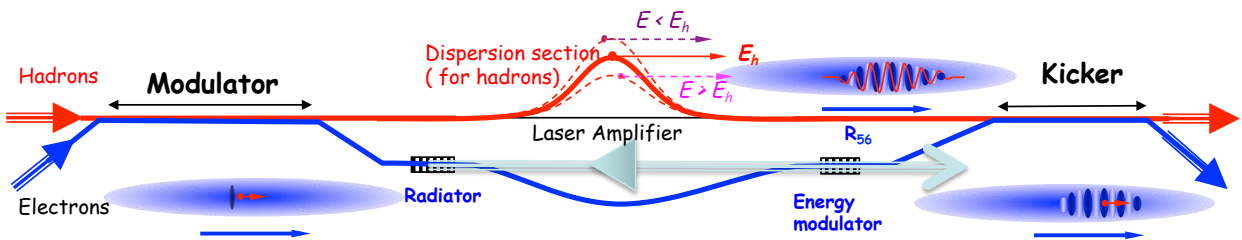

Figure 1.c. A CeC schematic [15] with hybrid laser-beam amplifier (HLBA). HLBA uses a broad-band laser amplifying electron-beam's radiation from a short wiggler. In a second wiggler the amplified laser power modulates the electrons energy. The latter is transferred into a density modulation using the $R_{56}$ of an achromatic dogleg.

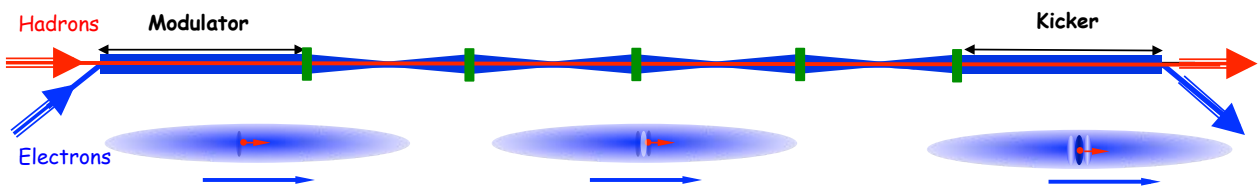

Figure 1.d. A layout of a CeC with a PCA. No beam's separation is necessary.

Cooling in transverse directions can be accomplished by coupling longitudinal and transverse degrees of freedom [12,15]. The electric fields induced by surrounding hadrons (as well as by shot noise in electron beam) would provide random kicks to the hadrons resulting in diffusion. Balance between the cooling and the diffusion would determine the final state of the hadron beam and its emittances.

In detail, the CeC mechanism is based on each hadron interacting in the kicker with self-induced and amplified density modulation in the electron beam. In order for this to happen the propagation time between the modulator and the kicker for the hadron with nominal energy (1) should be equal to that of the high density slice. If the group velocity of the e-beam instability (e.g. that of the wave-packet or spike of the density modulation) is equal or slightly higher than that of the hadron beam, both beams can co-propagate through system without being separated. If necessary, an achromatic chicane installed onto the common path could introduce delay of the electron beam with respect to much heavier hadrons.

But most of the schemes shown in Fig.1 do not satisfy this requirement. As shown in [20] the group velocity of the wave-packet in high-gain FEL amplifier is given by following expression

$$\frac{v_g - v_o}{c} \cong \frac{\alpha - (1-\alpha)a_w^2}{2\gamma_o^2}, \; for \; \gamma_o^2 \gg 1 \qquad (2)$$

where $a_w = eB_w\lambda_w/m_ec^2$ is dimensionless parameter of the FEL wiggler with period $\lambda_w$ and peak magnetic field $B_w$[1]. Coefficient $\alpha$ depends on the diffraction of FEL light and theoretically has maximum value of 1/3 for a 1D FEL [21]. It is typically below 1/4 for a realistic 3D FEL amplifier [20]. Requirement $v_g \geq v_o$ therefore limits wiggler parameter to $a_w < \sqrt{\alpha/(1-\alpha)}$, which can limit FEL performance. This mode of FEL operation with $a_w$=0.5 is used for our CeC proof-of-principle experiment, which is in progress at Relativistic Heavy Ion Collider (RHIC), BNL [27-29]. Using an FEL amplifier with larger $a_w$ would require separating electron and hadron beams and bending hadron beam trajectory.

In CeC based on MBIA and HLBA (Figs. 1 b-c) the electron beam is always delayed with respect to the hadron beam, and consequently the density imprint of the hadron is delayed. These schemes would always require delaying the hadron beam, which is only possible when much heavier hadrons are separated from electron beam. This CeC mode is illustrated in Figs. 1 a-c with hadron bypasses. This part of the CeC system would be extremely expensive, and if possible, should be avoided.

This was the main reason for our searching of electron beam instability as broadband as MBIA, but occurring without delay of electron beam. This search, started from considering a possibility of parametric amplifier suggested in original CeC papers by Derbenev [18-19], brought us to fast-growing plasma instability we call Plasma-Cascade Amplifier, or PCA.

---

[1] Here we are using $a_w$ defined for a helical wiggler.

## II. Plasma-Cascade Amplifier – principle of operation

The idea of the PCA is both very simple and counter-intuitive. It is counter-intuitive because the PCA micro-bunching instability uses, generally speaking, a repulsive space charge forces for over-focusing of longitudinal plasma oscillations.

The exponentially growing "over-focusing" instability, resulting in unstable ray trajectory, is a well-known phenomenon in periodic focusing optical system– see Fig. 2. The instability occurs when the focal length of its lenses, $f$, is less than a quarter of distance between them, $L=2l$, e.g.:

$$l > 2f. \tag{3}$$

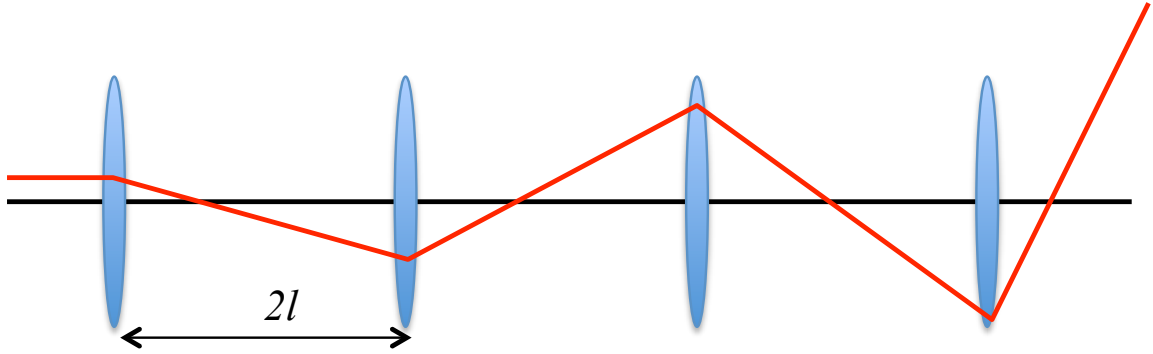

Figure 2. Unstable ray trajectory in a system of periodic focusing lenses with focal length $f < l/2$.

In beam optics terms, the instability occurs when the transport matrix of the periodic cell has an eigen value with modular more than unity - in this case $\lambda_1 < -1$:

$$\mathbf{M}_c = \begin{bmatrix} m_{11} & m_{12} \\ m_{21} & m_{22} \end{bmatrix} = \begin{bmatrix} 1 & l \\ 0 & 1 \end{bmatrix}\begin{bmatrix} 1 & 0 \\ -f^{-1} & 1 \end{bmatrix}\begin{bmatrix} 1 & l \\ 0 & 1 \end{bmatrix} = \begin{bmatrix} 1-lf^{-1} & l(2-lf^{-1}) \\ -f^{-1} & 1-lf^{-1} \end{bmatrix}; \tag{4}$$

$$\lambda_{1,2} = m_{11} \mp \sqrt{m_{11}^2 - 1};$$

For unstable motion (3) both eigen values of symplectic transport matrix (e.g. $\det \mathbf{M}_c = 1$, $\lambda_1 \lambda_2 = 1$) are real:

$$lf^{-1} > 2 \rightarrow m_{11} = 1 - lf^{-1} < -1; \quad m_{12} = l(2-lf^{-1}) < 0;$$
$$\lambda_1 = m_{11} - \sqrt{m_{11}^2 - 1} < -1; \quad \lambda_2 \equiv \lambda_1^{-1}. \tag{5}$$

Eigen vectors $Y_{1,2}$ for unstable motion can be easily found

$$\mathbf{M}_c Y_{1,2} = \lambda_{1,2} Y_{1,2}; \quad Y_{1,2} = \begin{bmatrix} w \\ \pm w^{-1} \end{bmatrix}; \quad w^2 = \frac{-m_{12}}{\sqrt{m_{11}^2 - 1}} > 0; \tag{6}$$

where we symplectically normalized them

$$Y_2^T \mathbf{S}_{(2d)} Y_1 \equiv -Y_1^T \mathbf{S}_{(2d)} Y_2 = 2; \quad \mathbf{S}_{(2d)} = \begin{bmatrix} 0 & 1 \\ -1 & 0 \end{bmatrix}. \tag{7}$$

The obvious condition $Y_1^T \mathbf{S}_{(2d)} Y_1 = Y_2^T \mathbf{S}_{(2d)} Y_2 = 0$ can be used in combination with (7) to derive the solution for the trajectory. The solution, after passing $n$ cells, reads:

$$X(n) = a_1 \lambda_1^n Y_1 + a_2 \lambda_1^{-n} Y_2;\ a_1 = \frac{1}{2} Y_2^T \mathbf{S}_{(2d)} X_o; a_2 = -\frac{1}{2} Y_1^T \mathbf{S}_{(2d)} X_o; X_o = \begin{bmatrix} x_o \\ x'_o \end{bmatrix};$$

$$X(n) = \begin{bmatrix} \dfrac{\lambda_1^n + \lambda_1^{-n}}{2} \\ \dfrac{\lambda_1^n - \lambda_1^{-n}}{2w^2} \end{bmatrix} x_o + \begin{bmatrix} w^2 \dfrac{\lambda_1^n - \lambda_1^{-n}}{2} \\ \dfrac{\lambda_1^n + \lambda_1^{-n}}{2} \end{bmatrix} x'_o. \qquad (8)$$

where $X_o$ corresponds to initial condition of the trajectory (ray).

It is known that evolution of small electron density perturbations $\tilde{n}(\vec{r})$ in cold, infinite and homogeneous plasma is described by plasma oscillations [25]:

$$\frac{d^2 \tilde{n}}{dt^2} + \omega_p^2 \tilde{n} = 0;\ \omega_p^2 = \frac{4\pi n_o e^2}{m}; \qquad (9)$$

where $n_o$ is the unperturbed electron's density. Thus, if we manage to periodically modulate the plasma frequency, for example by modulating the electron density $n_o \to n_o(t)$, we may reach similar over-focusing instability. The most natural way of modulating the beam density[2] is using transverse focusing – an example of a periodic transverse focusing and modulating beam's transverse size using solenoids is shown in Fig. 3.

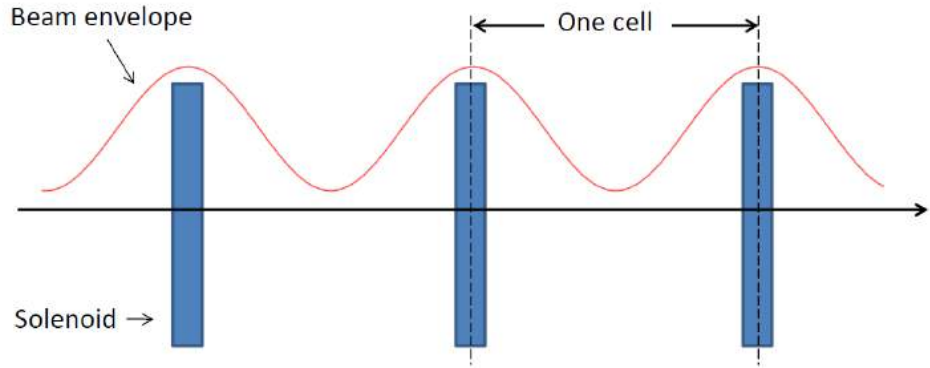

Figure 3. Illustration of the beam envelope evolution along a periodic transport system with solenoids as focusing elements.

---

[2] While periodic bunching and de-bunching of electron bunch is theoretically possible, it is definitely much more complicated, if even practical. It would require multiple RF cavities for chirping and de-chirping beam energy as well as large $R_{56}$ of transports lines between the cavities. The later may require bending electron beam trajectory, which defies the entire purpose of our approach to PCA.

A self-consistent solution of the beam envelope, which we discuss in next section, defines the beam waist size, $a_o$, in the middle of each cell as well as evolution of the envelope $a$ inside the cell. Strength of solenoid focusing is selected to support this periodic beam envelope.

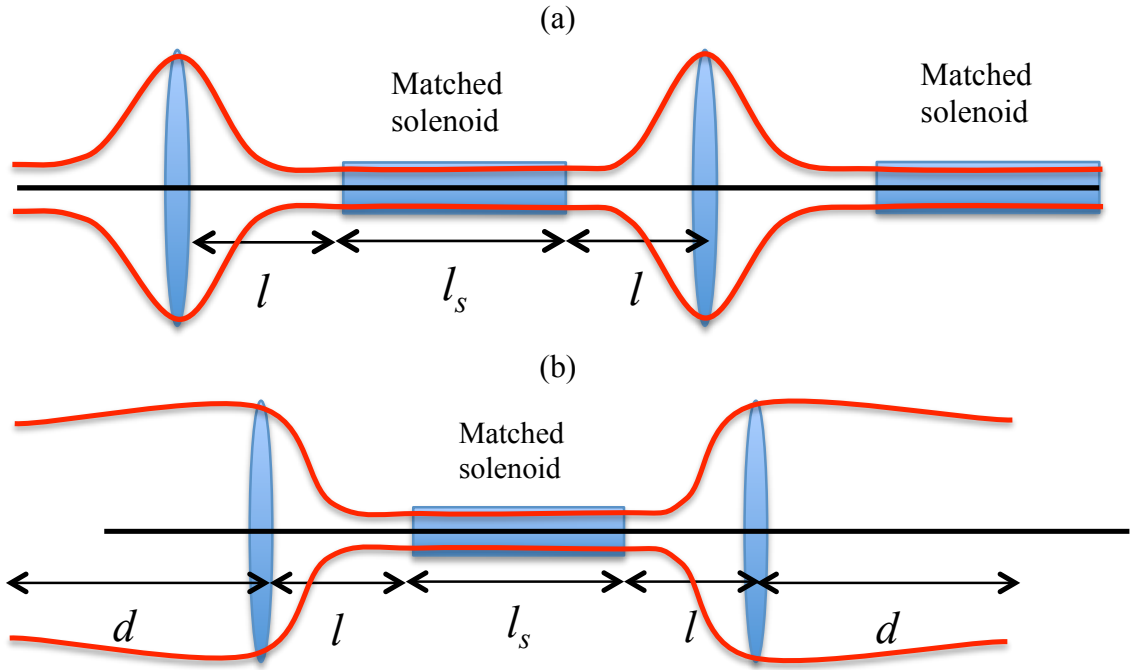

Figure 4. Alternative schemes providing for separation of the "strong focusing" in the cell center and "fast expansion" outside it (a) and also for reducing cell's chromaticity (b) by early interspersing the beam size expansion.

Figs. 4(a) and 4(b) alternative cell schemes for PCA, which allow additional control of the processes by extending waist of the beam by matching it into a long solenoid. Scheme shown in Fig. 4(b) would also allow controlling chromaticity of such strong focusing cell. Having control of additional parameter, as we discuss later, will definitely allow improving PCA performance. While considering schemes shown in Fig. 4 as possibility for future optimization of the PCA, in this paper we are considering in detail scheme shown in Fig. 3.

Modulation of the electron density is inverse proportional to the square of the beam radius $a$, which could vary significantly inside the cell (see details in next section) making plasma oscillations (9) unstable. The question remains: what conditions (e.g. what electron beam parameters) are required for this instability to occur and what would be its growth rate?

In following sections we give analytical description of this process to address the above questions. We then examine assumptions used in our studies and limitation of the method. It followed by results of self-consistent 3D computer simulations, showing that PCA is indeed can work and provide very significant gain in the longitudinal density modulation. Finally, we discuss the proposed experimental demonstration of PCA-based CeC using existing system we had built at RHIC. We conclude this paper with discussion of the applicability of this CeC scheme for cooling hadron beam in future EIC.

## III. Analytical description

The goal of this analytical description of PCA is to find a self-consistent model, which
(a) would allow analytical description of the problem and
(b) define set of dimensionless parameters of the problem.

Naturally this would require some assumptions and approximations. First, we would describe how we decouple transverse and longitudinal degrees of freedom. Then we would derive - and when possible solve - dimensionless differential equations for transverse and longitudinal distribution functions.

In this section, we firstly show that up to the leading order of $(x, p_x, y, p_y, t, E)$, the solution of the Vlasov equation for the electrons can be factorized into the longitudinal and transverse parts. By assuming K-V distribution for the 4-D transverse space, the transverse part of the solution can be readily found, which provides information about how transverse beam size, and hence the spatial density of electrons, evolves with time. We then show that, for uniform distribution of the line charge density and κ-1 (Lorentzian) longitudinal momentum distribution, the longitudinal part of the Vlasov equation can be linearized and re-written into an O.D.E. of the second order. The homogeneous part of the O.D.E. has the form of the Hill's equation, which we are solving both analytically and numerically. We are interested in the unstable solutions and we are investigating the growth rate per cell dependence on two parameters representing the space charge forces and the transverse beam dynamics.

### III.1. Decoupling Vlasov Equations

The 6D Vlasov equation for an electron beam in an accelerator can be written as

$$\frac{\partial}{\partial s} f(s, X) + \frac{\partial f}{\partial X} \cdot \left[ S \frac{\partial}{\partial X} H(s, X) \right] = 0 ,  \quad (10)$$

where $s$ is the longitudinal coordinate along the accelerator and

$$X^T = (x, P_x, y, P_y, -t, E) \text{ or } (x, x', y, y', \tau, \delta);$$

$$\tau = c\left(\frac{s}{v_o} - t\right); \; \delta = \frac{E - E_o}{p_o c}; \; p_o c = \sqrt{E_o^2 - m_e^2 c^4} = \beta_o E_o; \beta_o = \frac{v_o}{c}, \quad (11)$$

is a 6-vector comprised of 3 Canonical pairs $(x, P_x)(y, P_y)(-t, E)$ or $(x, x')(y, y')(\tau, \delta)$. The system's Hamiltonian $H(s, X)$ [26] governs the particle's motion:

$$\frac{d}{ds} X = S \frac{\partial}{\partial X} H(s, X) , \quad (12)$$

with a generator of symplectic group $S$ - a 6x6 block diagonal matrix:

$$S = \begin{bmatrix} \sigma & 0 & 0 \\ 0 & \sigma & 0 \\ 0 & 0 & \sigma \end{bmatrix}; \sigma = \begin{bmatrix} 0 & 1 \\ -1 & 0 \end{bmatrix}. . \quad (13)$$

In the absence of coupling between longitudinal direction and transverse directions the Hamiltonian can separated into two independent parts:

$$H(s,X) = H_\perp(s,X_\perp) + H_{//}(s,X_{//}), \tag{14}$$

where

$$X_\perp = (x,p_x,y,p_y) \text{ or } (x,x',y,y') \tag{15}$$

are a particle's coordinates in 4-D transverse phase space and

$$X_{//} = (-t,\mathbf{E}) \text{ or } (\tau,\delta); \tag{16}$$

are a particle's coordinates in the 2-D longitudinal phase space [27]. It follows from eq. (14) that the second term from the L.H.S. of eq. (10) reduces to

$$\frac{\partial f}{\partial X} \cdot \left[ S \frac{\partial}{\partial X} H(s,X) \right] = \frac{\partial f}{\partial X_\perp} \cdot \left[ S^{(4d)} \frac{\partial}{\partial X_\perp} H_\perp(s,X_\perp) \right] + \frac{\partial f}{\partial X_{//}} \cdot \left[ S^{(2d)} \frac{\partial}{\partial X_{//}} H_{//}(s,X_{//}) \right], \tag{17}$$

with

$$S^{(4d)} = \begin{bmatrix} \sigma & 0 \\ 0 & \sigma \end{bmatrix}, \tag{18}$$

and

$$S^{(2d)} = \sigma = \begin{pmatrix} 0 & -1 \\ 1 & 0 \end{pmatrix}. \tag{19}$$

Assuming that the distribution function can be factorized

$$f(s,X_\perp,X_{//}) = f_\perp(s,X_\perp) f_{//}(s,X_{//}), \tag{20}$$

and inserting eq. (17) and eq. (20) into eq. (10) yields:

$$f_{//}^{-1}(s,X_{//})\left\{ \frac{\partial}{\partial s} f_{//}(s,X_{//}) + \frac{\partial f_{//}(s,X_{//})}{\partial X_{//}} \cdot \left[ S^{(2d)} \frac{\partial}{\partial X_{//}} H_{//}(s,X_{//}) \right] \right\}$$
$$+ f_\perp^{-1}(s,X_\perp)\left\{ \frac{\partial}{\partial s} f_\perp(s,X_\perp) + \frac{\partial f_\perp(s,X_\perp)}{\partial X_\perp} \cdot \left[ S^{(4d)} \frac{\partial}{\partial X_\perp} H_\perp(s,X_\perp) \right] \right\} = 0 \tag{21}$$

In order to satisfy eq. (21) for arbitrary $X_\perp$ and $X_{//}$, both terms inside the figure brackets have to vanish, which leads to the following decoupled Vlasov equations:

$$\frac{\partial}{\partial s} f_{//}(s,X_{//}) + \frac{\partial f_{//}(s,X_{//})}{\partial X_{//}} \cdot \left[ S^{(2d)} \frac{\partial}{\partial X_{//}} H_{//}(s,X_{//}) \right] = 0, \tag{22}$$

and

$$\frac{\partial}{\partial s} f_\perp(s,X_\perp) + \frac{\partial f_\perp(s,X_\perp)}{\partial X_\perp} \cdot \left[ S^{(4d)} \frac{\partial}{\partial X_\perp} H_\perp(s,X_\perp) \right] = 0. \tag{23}$$

### III.2. Transverse motion

First let's solve the transverse part of the Vlasov equation and find how transverse beam size, *a*, evolves along the beamline. In other words, we are finding solution for the envelope equation [28]. Let's consider a beamline with a set of periodic cells consisting of a short focusing element (for example a solenoid) and a drift as shown in Fig. 2.

Lets' consider a round beam with constant current $I_o$ and radius *a*, which is function of the azimuth *s*:

$$j_o = en_o v = \frac{I_o}{\pi a^2}; \quad n_o = \frac{I_o}{ev_o} \cdot \frac{1}{\pi a^2} \tag{24}$$

It is known that analytical self-consistent solution for a continuous beam[3] is possible for K-V distribution of electrons in 4-D transverse phase space [1]:

$$f_\perp(x, y, x', y') = f_o \delta(I_x + I_y - \varepsilon) \tag{25}$$

where $I_{x,y}$ are actions of transverse motions (Courant-Snider invariants [29])

$$I_x = \frac{x^2}{w^2} + (wx' - w'x)^2 = inv; \quad I_y = \frac{y^2}{w^2} + (wy' - w'y)^2 = inv; \tag{26}$$

and $\varepsilon$ is so called geometrical emittance of the beam envelope and

$$\beta \equiv w^2; \quad a = \sqrt{\varepsilon \beta} \tag{27}$$

is β-function of transverse (betatron) motions [29]. The resulting transverse distribution is an evenly charged cylinder with radius *a*

$$\rho(x, y) = \iint f_\perp dx' dy' = \begin{cases} \frac{\pi f_o}{\beta}, r^2 \le a^2 \\ 0, r^2 > a^2 \end{cases}, \tag{28}$$

with a linearly depending radial electric field inside the beam envelope [28]. The resulting second order nonlinear ordinary differential equation for beam envelope $a(s)$ reads (see eq. (4.112) in ref [28]):

$$a'' + K(s)a - \frac{2}{\beta_o^3 \gamma_o^3} \frac{I_o}{I_A} \frac{1}{a} - \frac{\varepsilon^2}{a^3} = 0; \quad K(s) = \left(\frac{eB_{sol}(s)}{2p_o c}\right)^2; I_A = \frac{mc^3}{e} \approx 17 \, kA, . \tag{29}$$

where $K(s)$ is the focusing strength of solenoids and $I_A = 4\pi\varepsilon_0 \frac{mc^3}{e} \approx 17 \, kA$ is the Alfven current. Any solution of (29) satisfies the transverse Vlasov equation (23).

---

[3] For a transversely uniform round beam with radius $a(s)$ the transverse field is proportional to local density when $\frac{da}{ds} \ll \gamma$, the condition which is easy to be practically satisfied.

We used the following strategy in determining the lattice (e.g. structure) of the cell, shown in Fig. 5.:

(a) We use bilateral symmetry for the cell with the beam waist with radius $a_o$ located in the middle of the cell;
(b) We solve self-consistently the envelope equation (29) in a drift space, $K = 0$, starting with initial conditions $\{a = a_o, a' = 0\}$ at $s=0$;
(c) We choose strength of solenoid in such a way that beam envelope reaches its maximum, $a' = 0$, in the center of the solenoid. This step is done numerically.

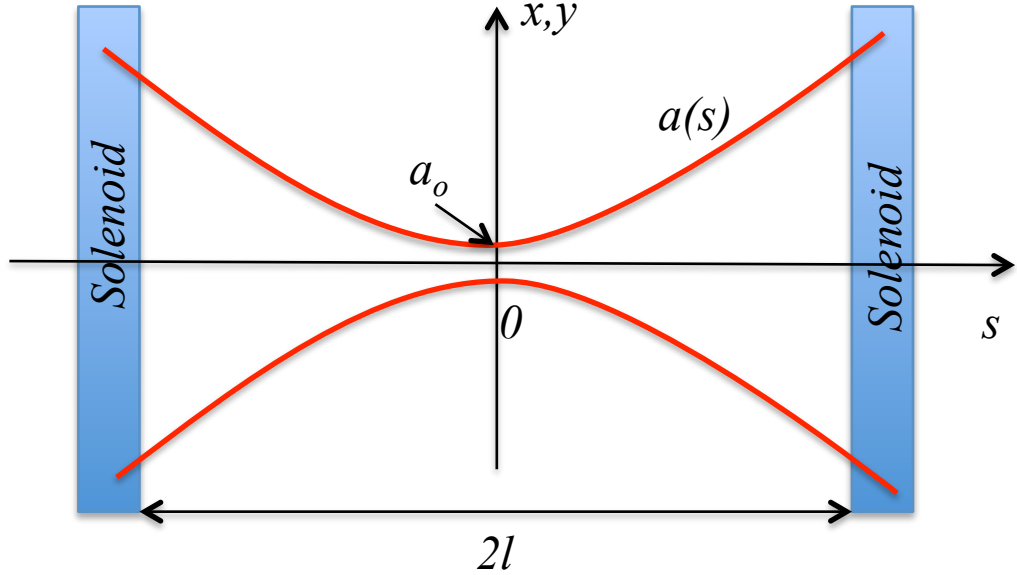

Figure 5. Shape of the beam envelope in a bilaterally symmetric cell with drift length of *2l* and short focusing solenoids.

This approach reduces our analytical studies to envelope equation in drift space, where eq. (29) can be reduced to a dimensionless form at an unit interval:

$$\frac{d^2\hat{a}}{d\hat{s}^2} - k_{sc}^2 \hat{a}^{-1} - k_\beta^2 \hat{a}^{-3} = 0;$$

$$\hat{a} = \frac{a}{a_o} \geq 1;\ \hat{s} = \frac{s}{l} \in \{-1,1\};\ k_{sc} = \sqrt{\frac{2}{\beta_o^3 \gamma_o^3} \frac{I_o}{I_A} \frac{l^2}{a_o^2}};\ k_\beta = \frac{\varepsilon l}{a_o^2} = \frac{l}{\beta^*} = \hat{\beta}^{*1},$$

(30)

where we define the waist $\beta$-function [29] as

$$a_o^2 \equiv \varepsilon \beta^*.$$

Hence, the beam envelope inside the cell is fully determined by two dimensionless parameters: the space charge, $k_{sc}$, and the geometric $k_\beta$. Equation (30) is *s*-independent and has an associated conserved Hamiltonian, which can be solved for $\hat{a}'$:

$$h = \frac{\hat{a}'^2}{2} - k_{sc}^2 \ln \hat{a} + \frac{k_\beta^2}{2\hat{a}^2} = \frac{k_\beta^2}{2} = inv; \quad \hat{a}' \equiv \frac{d\hat{a}}{d\hat{s}} = \sqrt{k_\beta^2(1-\hat{a}^{-2}) + 2k_{sc}^2 \ln \hat{a}}. \quad (31)$$

It can be solved for two degenerated cases:

$$k_{sc} = 0, k_\beta \neq 0: \hat{a}^2 = 1 + \hat{s}^2/\hat{\beta}^2$$

and

$$k_{sc} \neq 0, k_\beta = 0 \to \hat{a} = Exp\left(\left\{Erfi^{-1}\left(\sqrt{\frac{2}{\pi}}k_{sc}\hat{s}\right)\right\}^2\right),$$

but its general solution cannot be expressed in terms of analytical function for $k_{sc} \neq 0, k_\beta \neq 0$.

Since equations of motion are Hamiltonian, we are solving eq. (30) using symplectic methods and testing accuracy of solutions with comparing the resulting Hamiltonian function with (30) and $\hat{a}'$ with (31). Our Hamiltonian has separate momentum and coordinates parts:

$$h = h_p + h_c; \{q = \hat{a}, p = \hat{a}'\}; h_p = \frac{p^2}{2}; h_c = -k_{sc}^2 \ln q + \frac{k_\beta^2}{2q^2}; \quad (31)$$

numerically solvable with the use the 4-th order symplectic integrator [30]:

$$\mathbf{M}(\xi) = \mathbf{M}_1\left(\frac{1}{2}A\xi\right)\mathbf{M}_2(A\xi)\mathbf{M}_1\left(\frac{\alpha}{2}A\xi\right)\mathbf{M}_2((\alpha-1)A\xi)\mathbf{M}_1\left(\frac{\alpha}{2}A\xi\right)\mathbf{M}_2(A\xi)\mathbf{M}_1\left(\frac{1}{2}A\xi\right);$$

$$\mathbf{M}_1(\xi) = \exp(-\xi : h_q :); \mathbf{M}_2(\xi) = \exp(-\xi : h_p :); \quad A = \frac{1}{1+\alpha}, \alpha = 1 - 2^{\frac{1}{3}}. \quad (32)$$

Using first-order maps for $\mathbf{M}_{1,2}$ and splitting interval $\hat{s} \in \{0,1\}$ into 200 steps reduces relative errors in $\hat{a}$ below $10^{-4}$ level for all range of parameters we used in these studies. Fig. 6 shows three examples of the beam envelope. We are using this detailed information for analytical studies of exponential growth (see next sections) arising from the Hill's-type *s*-dependent equation for the longitudinal density modulation. In addition, we used an approximate fitting of the beam envelope square into a parabola for analytical solution *s*-dependent of Hill's equation. We found that in range of interest (e.g. where the growth rates are close to maximum, an area around the line $k_{sc} \propto k_\beta/3$) the dimensionless envelope can be approximately described as

$$\hat{a}^2(\hat{s}) \approx 1 + (k_\beta^2 + 2k_{sc}^2)\hat{s}^2. \quad (33)$$

We should note that without space charge, $k_{sc} = 0$ e.q. (33) is an exact solution for the beam envelope corresponding to well known $\beta(s) = \beta^* + s^2/\beta^*$.

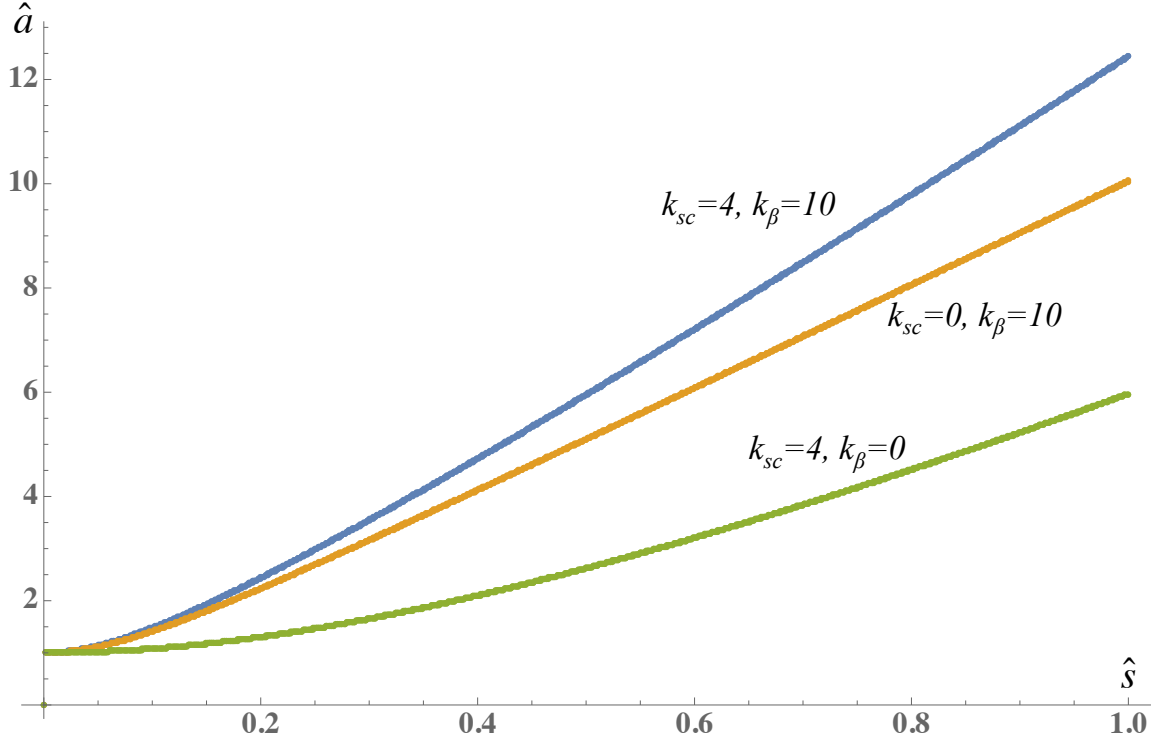

Figure 6. Calculated beam envelopes for three sets of parameters: green curve $k_{sc}=4, k_\beta=0$, yellow curve $k_{sc}=0, k_\beta=10$ and blue curve $k_{sc}=4, k_\beta=10$.

### III.3. Equation for Line Density Perturbation

While usually plasma oscillations in charged beams are described in a co-moving (beam) frame, here we will write the equation in the laboratory frame to avoid transformation back-and-forth from the frame to frame and $t$ to $s$ as independent variables. In the laboratory frame the longitudinal Vlasov equation also has a very simple form:

$$\frac{\partial}{\partial s} f_{//}(s,\delta,\tau) + \tau' \frac{\partial}{\partial \tau} f_{//}(s,\delta,\tau) + \delta' \frac{\partial}{\partial \delta} f_{//}(s,\delta,\tau) = 0;$$
$$\tau' = \frac{d\tau}{ds} = \frac{\delta}{\gamma_o^2 - 1} \equiv \frac{\delta}{\beta_o^2 \gamma_o^2}; \quad \delta' = \frac{d\delta}{ds} = -\frac{eE_s}{\gamma_o \beta_o m_e c^2};$$
(36)

For the next step in our analytical description we need to use an important approximation. Specifically we will use a high-frequency approximation for the longitudinal density modulation an an uniform transverse density. In other words, we assume that transverse expanse of the beam justifies use of 1D model for the electric field and the decoupling of the transverse and the longitudinal motions. Detailed consideration show that such approximation is valid only when the transverse extend of the beam is significantly larger than scale of the longitudinal modulation scaled by $\gamma_o$. When considering a longitudinal density variation $f_{//} \sim \exp(ik\tau)$, the above assumption is correct only if

$$ka >> \gamma_o.$$
(37)

In this case, for the core of the beam we can neglect transverse components of the electric field and the equation for the electric field is reduced to much simple differential equation:

$$\left|\frac{\partial E_{x,y}}{\partial x, y}\right| \ll \left|\frac{\partial E_s}{\partial s}\right| \Rightarrow \nabla \cdot \vec{E} \approx \left.\frac{\partial E_s}{\partial s}\right|_{t=const} = -4\pi e n, \tag{38}$$

where $E_s$ is the longitudinal electric field and $n$ is $r$-independent electron beam density. The later is the product of two components, the background e-beam density $n_o$ and the line density of the electron beam $\rho$:

$$n(s,\tau) = n_o(s) \cdot \rho(s,\tau); \quad n_o = \frac{I_o}{ev_o} \cdot \frac{1}{\pi a(s)^2}; \quad \rho(s,\tau) = \int_{-\infty}^{\infty} f_{//}(s,\delta,\tau) d\delta. \tag{39}$$

The unperturbed background line density is constant and is normalized to unity:

$$\rho_o = \int_{-\infty}^{\infty} f_o(\delta) d\delta = 1. \tag{40}$$

Assuming an infinitesimal perturbation $\tilde{f}$ of a distribution function:

$$f_{//}(s,\delta,\tau) = f_o(\delta) + \tilde{f}(s,\delta,\tau); \tag{41}$$

and noting that electric field is proportional to the perturbation:[4]

$$\frac{\partial E_s}{\partial s} = -4\pi e n_o(s) \cdot \tilde{\rho}(s,\tau); \quad \tilde{\rho}(s,\tau) = \int \tilde{f}(s,\delta,\tau) d\delta \tag{42}$$

we can neglect second order terms $E_s \tilde{f} \sim \tilde{f}^2$ in Vlasov equation

$$\frac{\partial \tilde{f}}{\partial s} + \frac{\delta}{\beta_o^2 \gamma_o^2} \cdot \frac{\partial \tilde{f}}{\partial \tau} - \frac{eE_s}{\gamma_o \beta_o m_e c^2}\left(\frac{df_0}{d\delta} + \frac{\partial \tilde{f}}{\partial \delta}\right) = 0; \tag{43}$$

---

[4] Since we are considering high frequency longitudinal beam density modulation, the space scale of modulation of interest is significantly smaller then the cell size $\delta s \propto a_o / \gamma_o \lll l$. In this case the density modulation and the corresponding electron can be factorized as a product of slow and fast varying functions.

$$\tilde{\rho}(s,\tau) = \rho_1(s) \cdot \rho_2(\tau); \left|\frac{d \ln \rho_1}{ds}\right| \ll \left|\frac{\partial \ln \rho_2}{\partial s}\right|_{t=const} = \beta_o^{-1}\left|\frac{\partial \ln \rho_2}{\partial \tau}\right|; \tau = \beta_o^{-1} s - ct;$$

$$E_s(s,\tau) = e_1(s) \cdot e_2(\tau); \left|\frac{d \ln e_1}{ds}\right| \ll \left|\frac{\partial \ln e_2}{\partial s}\right|_{t=const} = \beta_o^{-1}\left|\frac{\partial \ln e_2}{\partial \tau}\right|.$$

This allows to move slowly variable parts of the outside of the integrals when applying Fourier transform to eq. (38):

$$\int_{-\infty}^{\infty}\left(\left.\frac{\partial E_s}{\partial s}\right|_{t=const} - 4\pi e n(s,\tau)\right) e^{-ik\tau} d\tau \cong -ik E_k(s) - 4\pi e n_k(s) = 0.$$

We explicitly use it in eq. (45).

and arrive to a Linearized Vlasov equation. Using eq. (38) and applying Fourier transformation $\int g \exp(-ik\tau)d\tau$ to eqs. (43) and (37) we get to following equation

$$\frac{\partial \tilde{f}_k}{\partial s} + \frac{ik\delta}{\beta_o^2 \gamma_o^2} \cdot \tilde{f}_k - \frac{e\tilde{E}_k}{\gamma_o \beta_o m_e c^2} \frac{df_0}{d\delta} = 0; \quad \tilde{f}_k(s,\delta) = \int_{-\infty}^{\infty} \tilde{f}(s,\delta,\tau) e^{-ik\tau} d\tau; \qquad (44)$$

with specific expression for electric field (see footnote above):

$$\tilde{E}_k(s) = \int_{-\infty}^{\infty} E(s,\tau) e^{-ik\tau} d\tau = \frac{4\pi i}{k} e n_o(s) \tilde{\rho}_k(s); \quad \tilde{\rho}_k(s) = \int_{-\infty}^{\infty} \tilde{f}_k(s,\delta) d\delta. \qquad (45)$$

Multiplying both sides of eq. (44) by $\exp(i\alpha\delta s); \alpha = k/\beta_o^2 \gamma_o^2$ leads to

$$\frac{\partial}{\partial s}\left[e^{i\alpha\delta s} \tilde{f}_k(s,\delta)\right] = e^{i\alpha\delta s} \frac{e\tilde{E}_k}{\gamma_o \beta_o m_e c^2} \frac{df_0(\delta)}{d\delta}. \qquad (46)$$

Integrating this equation over $s$, and making use of eq. (45) yields

$$\tilde{f}_k(s,\delta) = e^{-i\alpha\delta s} \tilde{f}_k(0,\delta) + \frac{4\pi i}{k} \frac{e^2}{\gamma_o \beta_o m_e c^2} \int_0^s \tilde{\rho}_k(s_1) \cdot n_o(s_1) \frac{df_0(\delta)}{d\delta} e^{-i\alpha\delta(s-s_1)} ds_1, \qquad (47)$$

which, after being integrated over $\delta$, becomes

$$\tilde{\rho}_k(s) = \int_{-\infty}^{\infty} e^{-i\alpha\delta s} \tilde{f}_k(0,\delta) d\delta + \frac{4\pi e^2}{\gamma_o^3 \beta_o^3 m_e c^2} \int_0^s \tilde{\rho}_k(s_1) \cdot n_o(s_1)(s-s_1) ds_1 \left\{ \int_{-\infty}^{\infty} f_0(\delta) e^{-i\alpha\delta(s-s_1)} d\delta \right\}, \qquad (48)$$

where we assume $f_o(\delta)\big|_{\delta=\pm\infty} = 0$. For simplicity we will use a $\kappa$-$1$ energy distribution with relative energy spread of $\sigma_\delta$:

$$f_o(\delta) = \frac{1}{\pi \sigma_\delta} \frac{\sigma_\delta^2}{\sigma_\delta^2 + \delta^2}, \qquad (49)$$

which turns the figure bracket in eq. (48) into an exponential Landau damping term:

$$\int_{-\infty}^{\infty} f_o(\delta) e^{-i\alpha\delta(s-s_1)} d\delta = \exp\left(-\left|\frac{k\sigma_\delta}{\beta_o^2 \gamma_o^2}(s-s_1)\right|\right). \qquad (50)$$

Inserting eq. (50) into eq. (48) leads to

$$\tilde{\rho}_k(s) = \int_{-\infty}^{\infty} e^{-i\alpha\delta s} \tilde{f}_k(0,\delta) d\delta + \frac{4\pi e^2}{\gamma_o^3 \beta_o^3 m_e c^2} \int_0^s \tilde{\rho}_k(s_1) \cdot n_o(s_1) \exp\left(-\left|\frac{k\sigma_\delta}{\beta_o^2 \gamma_o^2}(s-s_1)\right|\right)(s-s_1) ds_1. \qquad (51)$$

Multiplying both sides of eq. (51) by $\exp\left(-(\beta_o \gamma_o)^{-2} |k\sigma_\delta (s-s_1)|\right)$ and then taking two successive $s$ derivatives of the resulting equation gives

$$\frac{d^2}{ds^2} \tilde{n}_k(s) + \frac{4\pi e^2 n_o(s)}{\gamma_o^3 \beta_o^3 m_e c^2} \tilde{n}_k(s) = -\frac{4\pi e^2 n_o(s)}{\gamma_o^3 \beta_o^3 m_e c^2} \int_{-\infty}^{\infty} e^{-i\alpha\delta s} e^{|\alpha|\sigma_\gamma s} \tilde{f}_k(0,\delta) d\delta, \qquad (52)$$

with

$$\tilde{n}_k(s) \equiv \tilde{\rho}_k(s)\exp\left(\left|\frac{k\sigma_\delta s}{\beta_o^2\gamma_o^2}\right|\right) - \int_{-\infty}^{\infty} e^{-i\frac{k\delta s}{\beta_o^2\gamma_o^2}} e^{\frac{|k\sigma_\delta s|}{\beta_o^2\gamma_o^2}} \tilde{f}_k(0,\delta)d\delta , \qquad (53)$$

Assuming the initial perturbation in the form of

$$f_1(0,\delta,\tau) = \rho_1(0,\tau) \cdot f_o(\delta) , \qquad (54)$$

with Fourier component of

$$\tilde{f}_k(0,\delta) = f_o(\delta)\tilde{\rho}_k(0) \qquad (55)$$

we obtain:

$$\int_{-\infty}^{\infty} e^{-i\frac{k\delta s}{\beta_o^2\gamma_o^2}} e^{\frac{|k\sigma_\delta s|}{\beta_o^2\gamma_o^2}} \tilde{f}_k(0,\delta)d\delta = \tilde{\rho}_k(0)\exp\left(-\left|\frac{k\sigma_\delta s}{\beta_o^2\gamma_o^2}\right|\right) . \qquad (56)$$

Inserting eq. (56) into eq. (52) produces

$$\frac{d^2}{ds^2}\tilde{n}_k(s) + \frac{4\pi e^2 n_o(s)}{\gamma_o^3\beta_o^3 m_e c^2}\tilde{n}_k(s) = -\frac{4\pi e^2 n_o(s)}{\gamma_o^3\beta_o^3 m_e c^2}\tilde{\rho}_k(0) , \qquad (57)$$

with

$$\tilde{n}_k(s) \equiv \tilde{\rho}_k(s)\exp\left(\left|\frac{k\sigma_\delta s}{\beta_o^2\gamma_o^2}\right|\right) - \tilde{\rho}_k(0). \qquad (58)$$

Eq. (57) and (58) can be rewritten in the form of Hill's equation:

$$\frac{d^2}{ds^2}\tilde{q}_k(s) + \frac{4\pi e^2 n_o(s)}{\gamma_o^3\beta_o^3 m_e c^2}\tilde{q}_k(s) = 0 , \qquad (59)$$

for new variable scaled up by Landau damping term:

$$\tilde{q}_k(k,t) \equiv \tilde{\rho}_k(s)\exp\left(\left|\frac{k\sigma_\delta s}{\beta_o^2\gamma_o^2}\right|\right). \qquad (60)$$

1D *s*-dependent Hill's equation is well known in accelerator physics with a straightforward ways of solving it and analyzing. Inside the cell, we can return to our dimensionless parameters rewriting (59) as:

$$\frac{d^2}{d\hat{s}^2}\tilde{q}_k + K_2(\hat{s})\cdot\tilde{q}_k = 0;\ K_2(\hat{s}) = 2\frac{k_{sc}^2}{\hat{a}(\hat{s})^2} . \qquad (61)$$

and solve it using standard methods developed for linear accelerator optics. First, we turn the second order differential equation into two first order differential equation in a matrix form:

$$X = \begin{bmatrix} \tilde{q}_k \\ \tilde{q}_k' = \dfrac{d\tilde{q}_k}{d\hat{s}} \end{bmatrix};\ \frac{d}{d\hat{s}}X = \mathbf{D}(s)X;\ \mathbf{D}(s) = \begin{bmatrix} 0 & 1 \\ -K_2(s) & 0 \end{bmatrix}, \qquad (62)$$

with transport matrix solution of

$$X(\hat{s}) = \mathbf{M}(0|\hat{s}) X(0); \quad \mathbf{M}(0|\hat{s}) = \exp\left(:\int_0^s \mathbf{D}(\xi)d\xi:\right) = \lim_{N\to\infty} \prod_{\substack{n=1 \\ \text{ordered}}}^{N} \mathbf{M}_n;$$

$$\prod_{\substack{n=1 \\ \text{ordered}}}^{N} \mathbf{M}_n \equiv \mathbf{M}_N \cdot \mathbf{M}_{N-1} \cdots \mathbf{M}_2 \cdot \mathbf{M}_1; \quad \mathbf{M}_n = \exp\left(\mathbf{D}(\hat{s}_n^*) \cdot \frac{\hat{s}}{N}\right), \hat{s}_n^* \in \left\{(n-1)\frac{\hat{s}}{N}, n\frac{\hat{s}}{N}\right\}; \quad (63)$$

$$\exp\left(\mathbf{D}(\hat{s}_n^*) \cdot \frac{\hat{s}}{N}\right) = \begin{bmatrix} \cos\varphi_n & \dfrac{\sin\varphi_n}{\omega_n} \\ -\omega_n \cdot \sin\varphi_n & \cos\varphi_n \end{bmatrix}; \quad \omega_n = \sqrt{K_2(\hat{s}_n^*)} = \sqrt{2}\frac{k_{sc}}{\hat{a}(\hat{s})}; \quad \varphi_n = \cdot\frac{\hat{s}}{N}.$$

Naturally we are interested in one cell symplectic matrix $\mathbf{M}_c = \mathbf{M}(-1|1)$ and its eigen values and vectors. Because of the bilateral symmetry $K_2(-\hat{s}) = K_2(\hat{s})$ the diagonal matrix elements $\mathbf{M}_c$ are equal [5]

$$\mathbf{M}_c = \begin{bmatrix} m_{11} & m_{12} \\ m_{21} & m_{11} \end{bmatrix}; \quad \det \mathbf{M}_c = 1. \quad (64)$$

"Trajectory" of plasma oscillations (61) is stable when $-2 < Tr(\mathbf{M}_c) < 2$ and is unstable when $|Tr(\mathbf{M}_c)| > 2$, e.g. $|m_{11}| > 1$. Solutions for unstable case, in which we are interested here, are identical to those we derived for focusing optics channel in eqs. (4), (6-8).

Since the evolution of $\hat{a}(\hat{s})$ inside the cell is fully determined by two parameters $k_{sc}$ and $k_\beta$

$$\hat{a}(\hat{s}) = F(\hat{s}, k_{sc}, k_\beta) \quad (65)$$

and the equation for transport matrix contains only $k_{sc}^2 / \hat{a}(\hat{s})^2$, one shall conclude that transport matrix of the cell as well as its eigen values are a functions of these two parameters:

$$\mathbf{M}_c = \mathbf{M}(-1|1) = \mathbf{M}(k_{sc}, k_\beta). \quad (66)$$

Hence, next section is dedicated to studying the dependence of cell's eigen values on $k_{sc}$ and $k_\beta$.

---

[5] It is well known how to relate symplectic matrices of a cell $\mathbf{M}$ and one of the cell's mirror image $\tilde{\mathbf{M}}$:

$$\mathbf{M} = \begin{bmatrix} a & b \\ c & d \end{bmatrix} \Rightarrow \tilde{\mathbf{M}} = \begin{bmatrix} d & b \\ c & a \end{bmatrix}; \mathbf{M}_c = \mathbf{M}\tilde{\mathbf{M}} = \begin{bmatrix} ad+bc & 2ab \\ 2dc & ad+bc \end{bmatrix}$$

## III.4. Growth rates of instability

To study PCA growth rate per-cell, we used direct evaluation of the cell transport matrix using a code written in Mathematica [31] and using its tools for graphical representation of the data. In addition, we used an approximate description for the envelope $\hat{a}(\hat{s})$ evolution (33) to solve Hill's equation (61) analytically:

$$\frac{d^2}{d\hat{s}^2}\tilde{q}_k + \frac{2k_{sc}^2}{1+\left(k_\beta^2+2k_{sc}^2\right)\hat{s}^2}\cdot\tilde{q}_k = 0. \tag{67}$$

We found that Hill's equation with inverse parabolic dependence of the $K_2$ strength has analytical solution

$$\frac{d^2x}{d\hat{s}^2} + \frac{\omega^2}{1+\kappa^2\hat{s}^2}\cdot x = 0;\quad d = \frac{\sqrt{\kappa^2-4\omega^2}}{4\kappa}$$

$$x = a_1 \cdot {}_2F_1\left(-\frac{1}{4}-d, -\frac{1}{4}+d, \frac{1}{2}, -\kappa^2\hat{s}^2\right) + a_2\cdot\kappa\hat{s}\cdot {}_2F_1\left(\frac{1}{4}-d, -\frac{1}{4}+d, \frac{3}{2}, -\kappa^2\hat{s}^2\right), \tag{68}$$

where $a_{1,2}$ are constants determined by initial conditions and

$${}_2F_1(a,b,c,z) \equiv F(a,b,c,z) = \frac{\Gamma(c)}{\Gamma(a)\Gamma(b)}\sum_{n=0}^{\infty}\frac{\Gamma(a+n)\Gamma(b+n)}{\Gamma(c+n)}\frac{z^n}{n!}$$

is hypergeometric function [32]. Inserting $\omega = \sqrt{2}k_{sc}$, $\kappa = \sqrt{k_\beta^2+2k_{sc}^2}$ into (68) gives analytical solution for eq. (67). This allows us to compare the exact analytical solution for an approximate $\hat{a}(\hat{s})$ with numerical solution for exact $\hat{a}(\hat{s})$. The comparison shown in Fig. 7 demonstrates reasonable agreement between these two models, especially in the area of fast growth (left-upper corner).

While the approximate analytical solution captures main features of exact numerical solution above the diagonal $k_\beta = k_{sc}$, it incorrectly predicts an addition area of stability at high values of $k_{sc}$. Hence, further in the paper we will refer to Fig. 7(b) as accurate description of PCA growth rates. It predicts two areas of stability – one next to the vertical axis where space charge forces are weak. Second stability line surrounding the line

$$k_\beta = \frac{4}{3}(k_{sc}-4)$$

can be explained by focusing at the ends of the cell, where deviation of "trajectory" is large, returns trajectory onto a stable path[6]. The rest of the right lower corner of the diagram has unstable motion with modest growth rates ($1 < \lambda < 5$). This mode is much harder to access in practice – it requires high peak current and low emittance beam – and represents mostly academic interest for PCA use for hadron cooling. Still, it can be important if one is studying micro-bunching instability in other schemes.

---

[6] In accelerator literature it is called an island of stability with tune advance per cell above 0.5 and below 1.

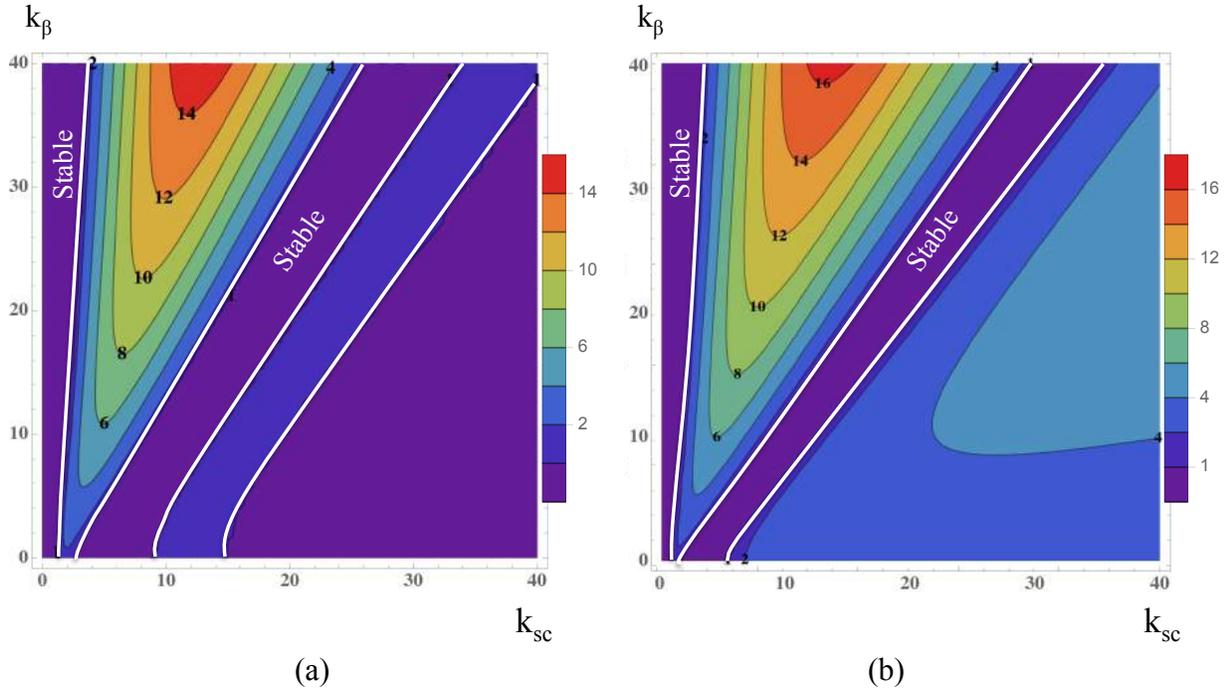

(a)                          (b)

Figure 7. Contour plots of $\lambda = \max(|\text{Re}\,\lambda_1|, |\text{Re}\,\lambda_2|)$, the absolute value of maximum growth rate per cell, using (a) an analytical solution (67) for an approximate beam envelope $\hat{a}(\hat{s})$ (as in eq. (33)), and (b) exact numerical solution of the problem using code described in Appendix A. Purple area highlighted by white lines indicates areas of stable oscillation $|\lambda_{1,2}| = 1$. Outside these areas oscillations are growing exponentially.

It worth noticing that space charge along cannot generate very large growth rates – even with $k_{sc} = 100$ (and $k_\beta = 0$) $\lambda$ reaching value ~ 3.6. Since required peak current is proportional for $k_{sc}^2$, it is very unlikely that this mode of PCA will be of practical interest for hadron cooling. At the same time, $\lambda = 4$ can be reached at modest parameters of $k_{sc} = 3$ and $k_\beta = 10$.

Fig. 8 shows 3D plot of exact solution for $\lambda = \max(|\text{Re}\,\lambda_1|, |\text{Re}\,\lambda_2|)$ using the same data as in Fig. 7(b) It clearly indicate that maximum growth rates are achieved along a "ridge" approximately following line $k_\beta = 3 \cdot (k_{sc} - 1.2)$. The ridge also can be seen in Fig. 7(b) as the line of maximum gradient.

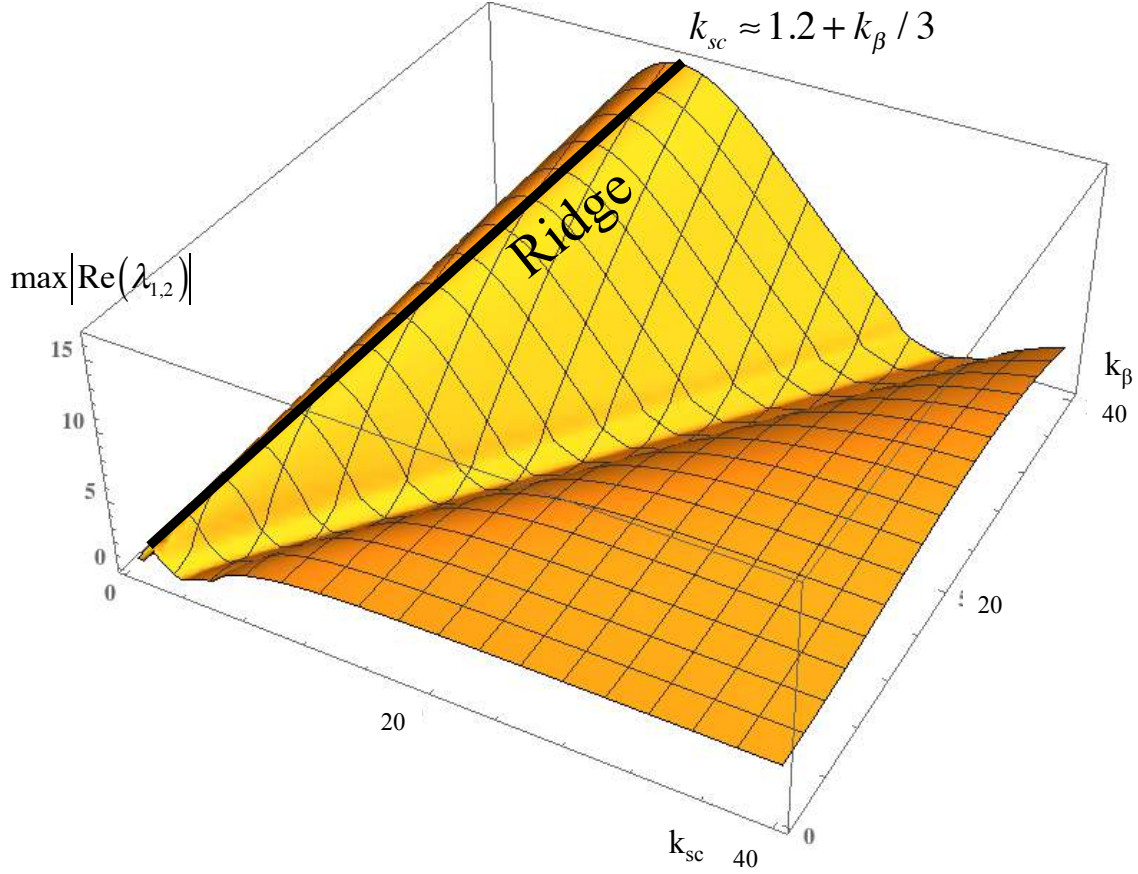

Figure 8. 3D plots of $\lambda = \max(|\text{Re}\,\lambda_1|, |\text{Re}\,\lambda_2|)$, the absolute value of maximum growth rate per cell, exact numerical solution. A ridge where growth rate peaks is clearly observed along the $k_\beta = 3 \cdot (k_{sc} - 1.2)$ line.

The area along the ridge is of the most interest for us and the growth on the ridge can be approximates by simple linear dependence:

$$\lambda \propto 1.25 k_{sc} \approx 1.5 + 0.413\, k_\beta. \tag{69}$$

A high gain PCA may require a number of cells and it is important to notice that eigen values are negative at and around the ridge. It means that for a positive amplification of the electron density, as required in a CeC, PCA should have even number of the cells: n=2m, $\lambda_1^{2n} > 0$. Furthermore, when applying exponential growth (8) with $\lambda^n \gg 1$:

$$\tilde{q}_k(n) = \frac{\lambda_1^n + \lambda_1^{-n}}{2}\tilde{q}_{ko} + w^2 \frac{\lambda_1^n - \lambda_1^{-n}}{2}\tilde{q}'_{ko} \to \frac{\lambda_1^n}{2}\tilde{q}_{ko} + w^2 \frac{\lambda_1^n}{2}\tilde{q}'_{ko} \tag{70}$$

one also should not forget about an additional factor 2 in the denominator, and that the expected amplification should be also reduced by $k$-dependent Landau damping (60):

$$\tilde{\rho}_k(n) \equiv \tilde{q}_k(n)\exp\left(-\left|\frac{2n \cdot kl}{\beta_o^2 \gamma_o^2}\sigma_\delta\right|\right). \tag{71}$$

The later, together with the time dependence of electrons arrival caused by angular spread

$$\delta t_{max} \approx \frac{nl}{c}\frac{\varepsilon_n}{\gamma_o \beta^*} \sim \frac{nl}{c}\frac{a_o^2}{\beta^{*2}} \quad (72)$$

will determine the PCA gain reduction at high frequencies (wave-numbers). At low frequency (wavenumbers) the PCA gain reduction will come from the

$$k < \frac{\gamma_o}{a_o}. \quad (73)$$

Further in the paper we will also use frequency, $f$ or/and $\omega$, for describing PCA based CeC, which is related to the wave-number $k$ via the beam velocity:

$$\omega \equiv 2\pi f = v_o k.$$

### III.4 Fundamental limitations of PCA gain and cooling rate.

The value of the PCA bandwidth is the most important parameter for other parameters of interest: its maximum gain and maximum cooling rate in PCA-based CeC.

First, the maximum cooling rate $\xi$ per turn of any stochastic system is determined by the longitudinal density of the beam(s) to be cooled and the frequency bandwidth of the pickup-amplifier-kicker system $\Delta f$ [9]. In our case both hadron and electron beams contribute into the noise in the CeC modulator, serving as the pick-up, and therefore increasing the size of the sample, $N_s$ [9], of the surrounding particles which affect an individual hadron

$$\xi_{max} \leq \frac{1}{N_s}, \quad N_s = N_{sh} + N_{se}; N_{sh} = \frac{1}{2\Delta f}\frac{I_h}{Ze}; N_{se} = \frac{1}{2\Delta f}\left|\frac{I_e}{e}\right|; \quad (74)$$

where $I_h$ and $I_e$ are peak current of hadron electron beams, correspondingly, and $Ze$ is electric charge of a hadron. Hence, maximum cooling rate of the hadron beam is fundamentally limited by the bandwidth of stochastic cooler. With typical proton peak currents ~ 10 A and modern collider circumferences ~ 10 km a bandwidth ~ 5 THz is needed to reach cooling rates below one hour. Modern RF-based stochastic cooling systems are short of this requirement by about three orders of magnitude. But as we can see from Table 1, PCA amplifiers - similarly to the high gain microbunching amplifiers proposed for CeC by Ratner [13] - can reach such bandwidth and go beyond.

By the nature of the micro-bunching amplifier the PCA bandwidth is in order of its highest frequency, which generally speaking, grows with beam energy. The scaling from equations (70-73) yields an approximation for frequency where PCA gain peaks:

$$f_{peak} \sim \frac{c}{4\pi L_{PCA}} \min\left(\frac{\beta_o^2 \gamma_o}{2\sigma_\delta}^2, \frac{\gamma_o \beta^*}{\varepsilon_n}\right). \quad (75)$$

We can reasonably estimate FWHM of the PCA amplifier is about half to one third of $f_{peak}$.

In addition to the bandwidth, any stochastic electron cooler for high energy hadrons should also have high gain, which has to be reached without saturating the amplifier [45-50]. Any CeC scheme requires its amplifier to operate on linear regime for resulting electric field to be a linear superposition of the fields induced by each hadron [12-14]. In other words, there should be no saturation in the amplifier originating either from hadron or electron beam. We had derived a very general estimate for maximum gain of the density modulation, which can be attained by instability in electron beam [46,48-50] and tested it against numerical simulations. According to our model-independent approach, the gain causing saturation in a longitudinal instability in electron beam depends of the peak beam current, central frequency of the instability and its bandwidth:

$$G_{sat} \approx \frac{1}{f_{peak}} \sqrt{\frac{I_e}{e} \Delta f} \;. \tag{76}$$

In the case of the two beams present in CeC, it should be reduced by additional factor $\sqrt{1 + Z \cdot I_h / I_e}$, which is insignificant in the cases listed below.

Table 1. Two examples of 4-cell PCA amplifiers for RHIC and eRHIC

| Parameter | CeC demonstration | RHIC/eRHIC |
|---|---|---|
| Relativistic factor of beams, $\gamma_o$ | 28.5 | 275 |
| e-beam peak current, $I_e$, A | 100 | 250 |
| Normalized transverse emittance, μm rad | 8 | 4 |
| Relative energy spread | $2 \cdot 10^{-4}$ | $10^{-4}$ |
| PCA cell length, m | 2 | 20 |
| e-beam radius at waist, mm | 0.2 | 0.05 |
| $\beta^*$, cm | 14 | 17 |
| $k_{sc}$ | 3.6 | 6.4 |
| $K_\beta$ | 7 | 29 |
| Estimated PCA gain | ~80 | ~120 |
| $f_{peak}$, Hz, estimation | $2.5 \cdot 10^{13}$ | $1.8 \cdot 10^{14}$ |
| $\Delta f$, Hz, estimation | $\sim 10^{13}$ | $\sim 10^{14}$ |
| $G_{sat}$ | ~ 3,000 | ~ 1,000 |

Table 1 clearly indicates that PCA, similarly to any broadband microbunching amplifier, can have very significant gain before while operating in linear regime.

**IV. Numerical simulations of PCA for CeC demonstration system**

While our PCA analysis uses rather sophisticated analytical model, it may still miss some important phenomena in real 3D beam dynamics. Hence, we conducted two sets of numerical studies to confirm that the PCA is a functioning model. For our numerical simulations we used two codes: SPACE [33] and PARMELA [34].

The code SPACE had been tuned for high fidelity simulations of the key processes in CeC [35-36] and includes all 3D beam dynamics necessary for PCA simulations. SPACE provides for generating initial beam distributions and propagating the beam through quadrupole or solenoid beam-line. It includes particle-in-cell (PIC) solvers for self-consistent EM fields with variety of boundary conditions, including periodic and open boundary conditions. In the PCA case electron's motion in the co-moving (beam) frame of reference is non-relativistic and we are using Poisson solver for this simulations. Since we are considering PCA with electron bunch whose length significantly exceed the scale of the density modulation, we are using periodic boundary conditions for a longitudinal bunch slice. In transverse directions we use open boundary condition to accurately simulate 3D effects imposed by the finite electron beam size.

The most interesting case that we simulated is the one we propose to test experimentally in the CeC system operating at RHIC facility [22-24]. As shown in Fig. 9, we propose to use existing 14.5 MeV CeC SRF electron accelerator and replace the central part of the CeC system (common with RHIC hadron beams) to a 4-cell PCA with five solenoids separated by 2 meters. The new section will serve as 4-cell PCA with microbunching gain about 100.

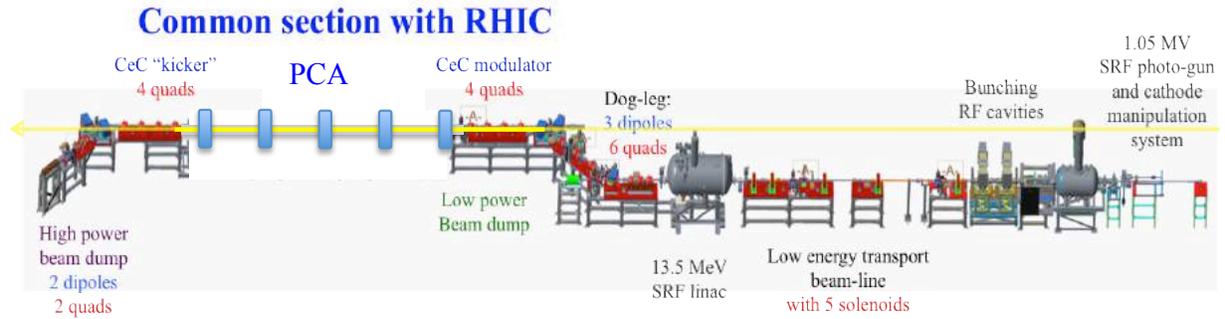

Figure 9. Proposed layout of CeC system at RHIC with new 8-m PCA section with 5 solenoids.

We also run start-to-end beam dynamics simulation in CeC accelerator using *Impact-T* code [37] confirming that beam parameters used for simulations can be attained in practice. Furthermore, similar parameters had been demonstrated experimentally in preparation for FEL based CeC demonstration experiment [23-24]. Finally, the beam envelope evolution in PCA was confirmed using code ASTRA [38].

First steps in SPACE PCA simulation were focused on developing self-consistent periodic beam envelope with desirable waist radius of 0.2 mm (see Fig. 10 (a)). It was followed by analysis of the evolution of the random longitudinal density modulation initiated by shot noise in electron beam. The shot noise is present both in velocity and density distribution, which causes a "ragged" gain's frequency dependence shown in Fig. 10 (b). A smooth dash-line shows expected response with high frequency reduction caused by the finite beam emittance and by the Landau damping (71-72). The most important observation from the spectral PCA response is that peak gain ~150 occurs at about 25 THz ($2.5 \cdot 10^{13}$ Hz), which is close to expected cut-off resulting from the energy and the angular spread in the beam (see Table 2). There is also clear low frequency PCA gain roll-off associated with the finite transverse beam size (73).

Table 2. Short list of main beam parameters in our simulations

| Parameter | PCA simulations | Start-to-end simulation* |
|---|---|---|
| e-beam energy, MeV | 14.564 | 14.564 |
| $\gamma_o$ | 28.5 | 28.5 |
| Peak current, A | 100 | 115 |
| Normalized transverse emittance, μm | 8 | 5.7 |
| Relative energy spread, RMS | $1 \cdot 10^{-4}$ | $< 2 \cdot 10^{-4}$ |
| Solenoid length, m | 0.2 | 0.2 |
| Solenoid field, kGs | 3.494 | 3.494 |
| Distance between solenoid, m | 2 | 2 |
| Transverse distribution | K-V | N/A |
| Beam radius in the waist, mm | 0.2 | 0.2 |
| $k_{sc}$ | 3.56 | ~4 |
| $k_\beta$ | 7.02 | ~10 |
| Energy spread frequency threshold, eq.(71), Hz | $2.4 \cdot 10^{13}$ | - |
| Emittance frequency threshold, eq.(72), Hz | $1.9 \cdot 10^{13}$ | - |
| Low frequency cut-off, eq. (73), Hz | $6.8 \cdot 10^{12}$ | - |

* These parameters are for 60% core of the 1 nC electron bunch.

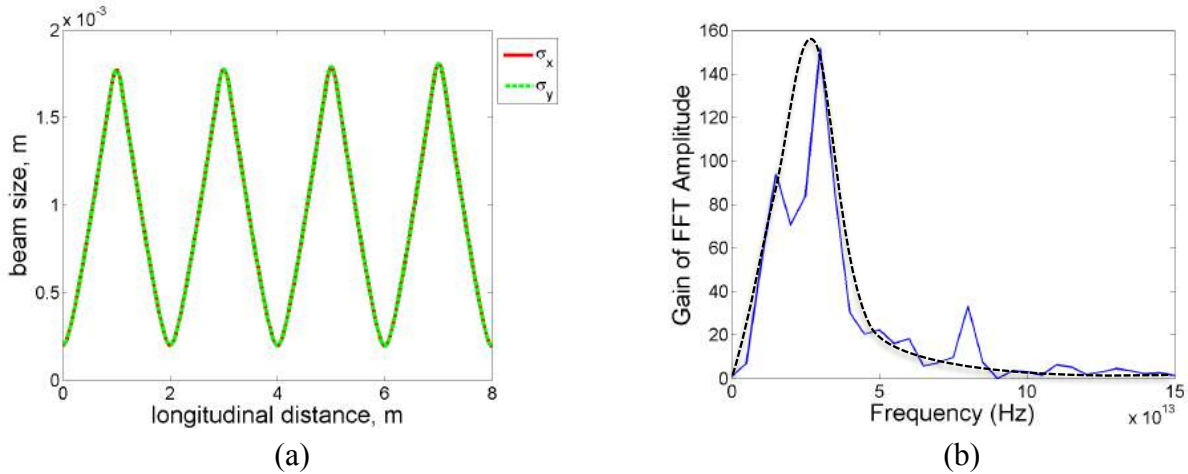

(a)        (b)

Figure 10. SPACE simulations results: (a) Self-consistent beam envelope in the PCA beam-line with 4 solenoids; (b) Change of the longitudinal density spectrum, starting from shot noise at *s=0* and evolving though 8-metter of PCA. Continuous line is the data and the dash-line is smooth curve representing expected gain.

A simple application of eq. (71), e.g. without taking into account Landau damping would indicate gain of 197. Applying Landau damping should bring the gain to about 117, which is lower than that in Fig. 10 (b). Hence, as we indicted above, a part of the amplified density modulation originates from the shot-noise in the initial velocity (energy) distribution.

To make a clear determination of the PCA gain we had performed SPACE simulations by introducing a small periodic density modulation at frequency of 25 THz, and propagated beam through the PCA. In order to eliminate background signal originated from the shot-noise, we used technique developed for high-fidelity CeC simulations [35-36]. We use two simulations with identical shot noise in the initial distribution of electron beam. First run is performed with short-noise only to define the background. The background as recorded as function of distance along the system. For the second simulation we add a small, but well-defined (for example periodic), density or energy modulation as initial signal. In absence of saturation, the amplified signal is obtained by subtracting the background. If necessary, we perform such simulation for a number of random seeds for the shot-noise.

Figure 11 shows evolution of longitudinal density modulation, originating from an initial 25 THz density modulation at $10^{-3}$ level, as the beam propagates through the 4-cell PCA. One can clearly see from Fig. 11 (b) an imprint of growing longitudinal plasma oscillations indicated by semi-periodic dips in the density modulations: at this locations the density modulation is transferred into the energy modulation, which then bounces back onto the density modulation. One also can see nearly exponential growth of the modulation from cell to cell. The amplification in the first cell is anticipated exception from pure exponential growth: here we can clearly see effect of factor 2 in the denominator in eq. (70). This "reduction" in the gain is nothing else than redistribution of initial modulation between two eigen modes: one with exponentially growing amplitudes as $\lambda_1^n$ and the other exponentially decaying as $\lambda_1^{-n}$.

As we expected the PCA gain reduces to about 80, which is in reasonable agreement with expectations that all three effects, the energy and angular spreads and the finite beam size, contribute into the gain reduction. In term of the "transport matrix" for density modulation, results presented in Fig. 11 represent quantitative evolution of its $m_{11}$ element. To evaluate quantitative evolution of "$m_{12}$ element of transport matrix", we simulated evolution of the density modulation originated from initial velocity modulation (in the beam frame) with results shown in Fig. 12.

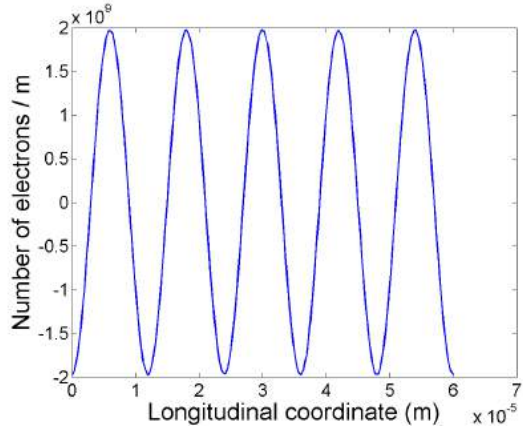
(a)

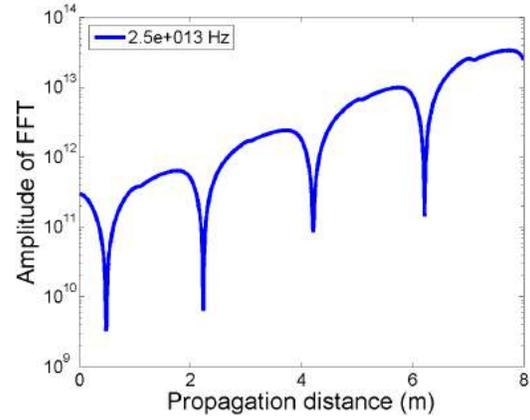
(b)

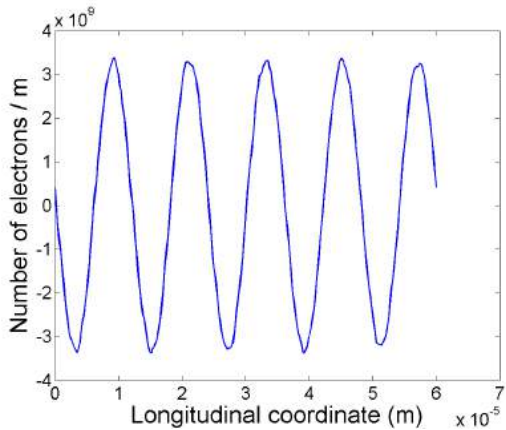
(c)

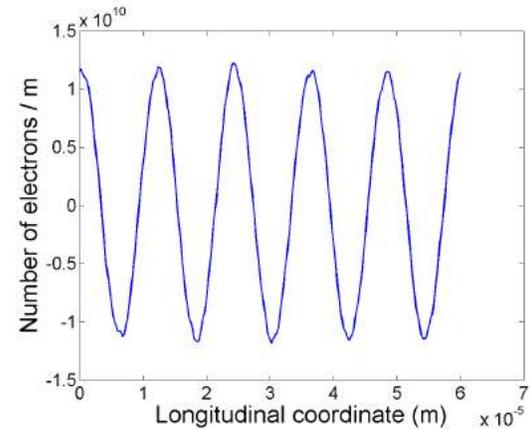
(d)

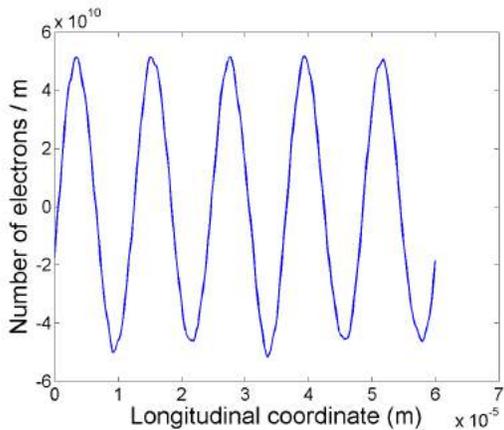
(e)

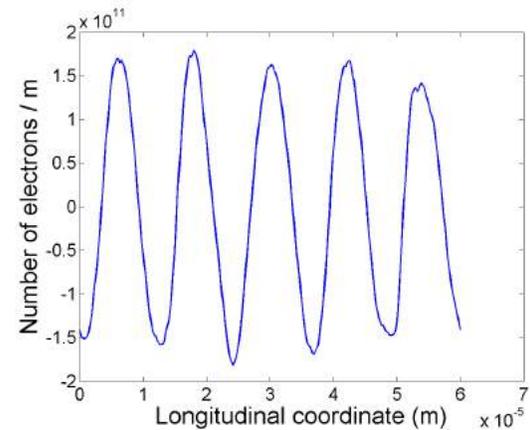
(f)

Figure 11. Evolution of longitudinal density modulation in the 4-cell PCA. Fig (a) shows initial density modulation at frequency of 25 THz and at $10^{-3}$ level of $2 \cdot 10^{12}$ m$^{-1}$ back-ground electron beam density. Fig. (b) shows the overall evolution of the corresponding Fourier component in the longitudinal density at 25 THz as function of the distance along the PCA. Longitudinal density modulation evolution: after one cell - (c), after two cells - (d), after three cells – (e), and after four cells - (f).

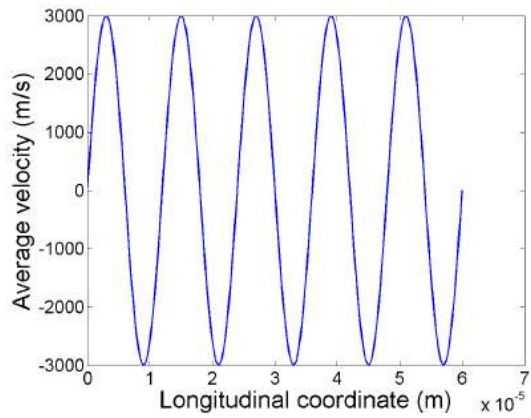

(a)

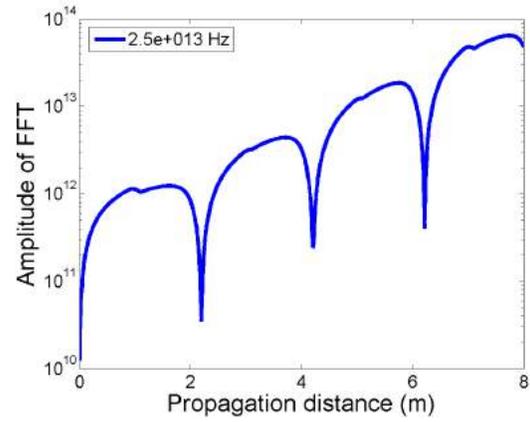

(b)

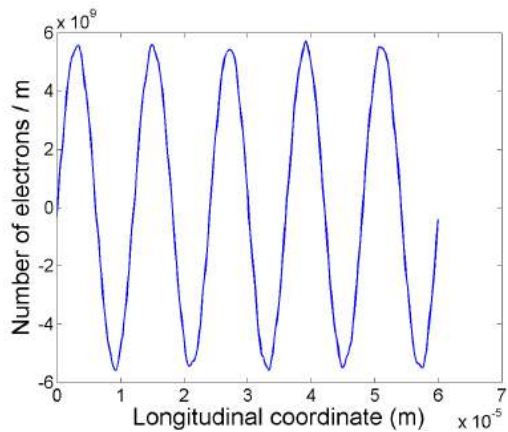

(c)

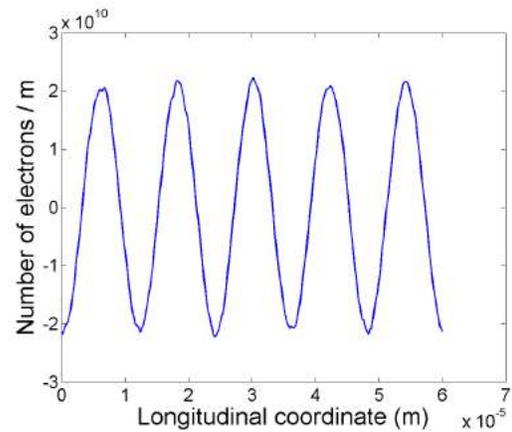

(d)

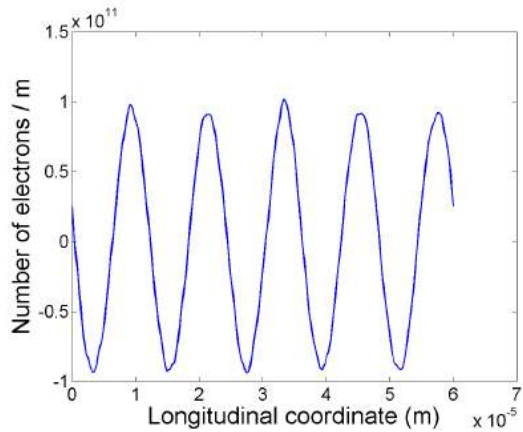

(e)

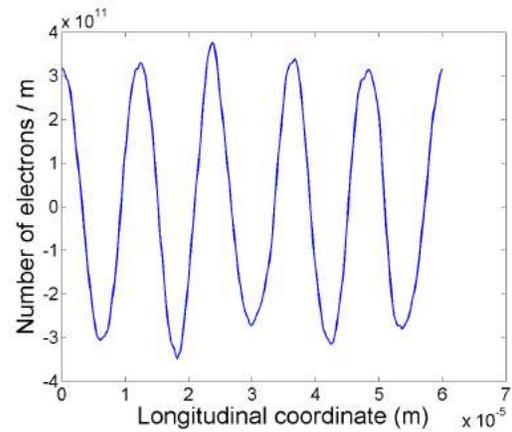

(f)

Figure 12. Evolution of longitudinal density modulation in the 4-cell PCA. Fig. (a) shows initial velocity modulation at frequency of 25 THz and amplitude of $10^{-5}c$ ($3 \cdot 10^3$ m/sec) in the beam frame (corresponding to energy modulation with relative amplitude of $10^{-5}$ in the lab frame). Fig. (b) shows the overall evolution of the 25 THz Fourier component in the longitudinal density at 25 THz as function of the distance along the PCA. Longitudinal density modulation evolution: after one cell - (c), after two cells - (d), after three cells – (e), and after four cells - (f).

From observing reasonable agreement between self-consistent 3D simulation and our analytical theory, which cannot accurately predict all 3D effects, we conclude that the PCA instability mechanism is robust against 3D effect.

Based on our understanding of the PCA amplification and its bandwidth, we made our standard estimations for the cooling rates for 26.7 GeV/u Au ions and protons using our CeC system with PCA amplifier. In recent years we had developed a number of analytical and numerical tools allowing us to estimate and to calculate CeC cooling rates [35-36,39-43]. First, we used our understanding of the density and energy modulation process in the CeC modulator [39,42] and our understanding of the gain and the frequency response of the PCA. We then simulated longitudinal time and space dependence of electric field in electron beam in the CeC kicker. Detailed considerations for main effects involved in this calculations are described in Appendices B and C.

Longitudinal electric field profile, shown in Fig. 13, is the result of initial density and energy modulation induced by a single $Au^{+79}$ ion in the CeC modulator section and amplified by the PCA. The parameters used in this plot are Au ions charge number $Z = 79$, plasma oscillation phase advance in the modulator $\psi_m = \pi/2$, electron beam radius in the kicker $a = 1.1 mm$, and relative energy spread in electron beam of $\sigma_\delta = 10^{-4}$ - see Appendix B for details.

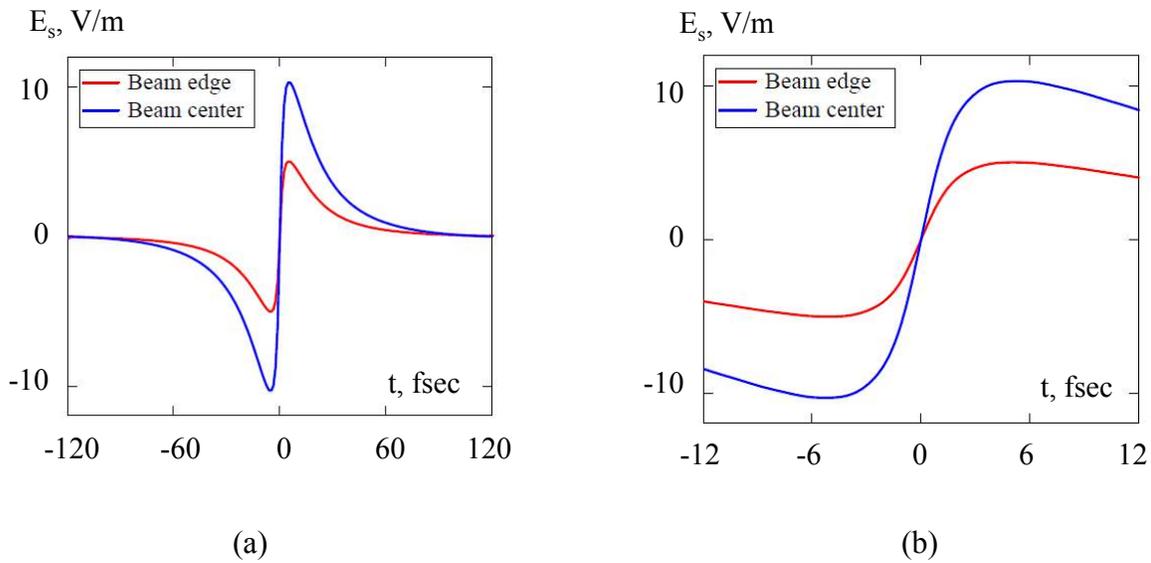

(a)        (b)

Figure 13. Time dependences of the longitudinal electric field in the CeC kicker. Horizontal axis is relative arrival time of hadron in femtoseconds (1 fsec $=10^{-15}$ sec). Fig. (b) has 1/10 of the time scale in fig. (a) to shows fine details of the field dependence. Vertical axis is longitudinal electric field in V/m. Blue line is the electric field in the beam center and red curve is that at the beam edge (e.g. at the radius of 1.1 mm).

As we described in the introduction, the condition (1) provide for ion with the nominal energy to arrive to the kicker at $t=0$, when the eclectic field is zero. More energetic ions arrive earlier and experience decelerating field. Visa versa, ions with lower energy are accelerated in the kicker. Maximum energy correction per turn of Au ion in the 3-meter long kicker is about 2.5 keV. Our estimations show that local cooling time for 26.7 GeV/u Au ion beam with RMS energy spread of 2·10⁻ circulating in RHIC with 3.8 km circumference would be about 6 seconds.

Cooling time for the whole bunch should be scaled by the ratio between lengths of the ion and the election bunches.

In our CeC demonstration experiments we use Au ion bunches with FWHM of 2 nsec and electron bunches with charge per bunch up to 4 nC. With 60% of useful electron beam having 100 A peak current the ratio between local and full-beam cooling time is ~ 80. Thus, we expect to have cooling time ~ 8 minutes.

We also propose to cool 26.7 GeV proton beam in RHIC using the same system, whose cooling time will be $Z^2/A$ ~ 32 times longer than that of Au ions. Hence, local cooling time for protons will be ~ 3 minutes, with full bunch cooling time ~ 4 hours. We hope to optimize the system further – for example by increasing gain of PCA to few hundreds - to improve on these parameters.

We also used code PARMELA for simulations this PCA and found a reasonable agreement between two codes.

## V. Numerical test of PCA with potential for low energy cooling in RHIC.

A low energy traditional electron cooling system, named LEReC [51], is under construction at RHIC for future low energy scan, motivated by the search of the QCD critical point. This system operates at relatively low electron energies of few MeV and, as shown in Fig. 14, contains large number of periodic sections with focusing solenoids.

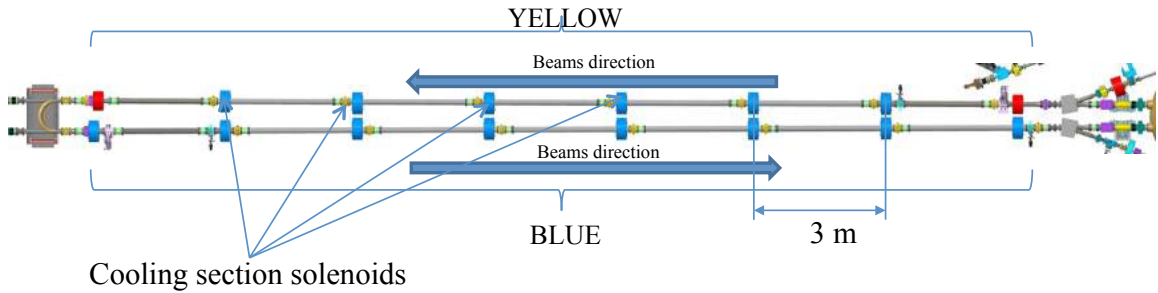

Fig. 14. Cooling section of LEReC system common with RHIC hadron beams (circulating in Yellow and Blue RHIC rings).

Favored scaling of PCA parameters with energy makes it attractive simulating this process in LEReC system. We used code PARMELA for these PCA with beam parameters listed in Table 3.

Simulation had been done for a 4 psec 0.2 pC longitudinal slice of the electron bunch. For PARMELA simulation we used initial energy modulation with amplitude of 50 eV at frequency of 4 THz. Gain exceeding ten was observed in case of 50 mA peak current. For peak current 100 mA we observed full saturation of the modulation as shown on Fig. 15.

While PCA with saturation cannot be used for cooling, these simulations are clear indication of susceptibility of low energy beams to plasma-cascade instability, which can be successfully used for broadband THz sources. It is also indication that this instability can play significant role in determining quality of the electron beam in long transport system such as LEReC.

We plan to continue studies of LEReC case using SPACE code to identify the range of parameters at which PCA and CeC can be tested at such very low energy.

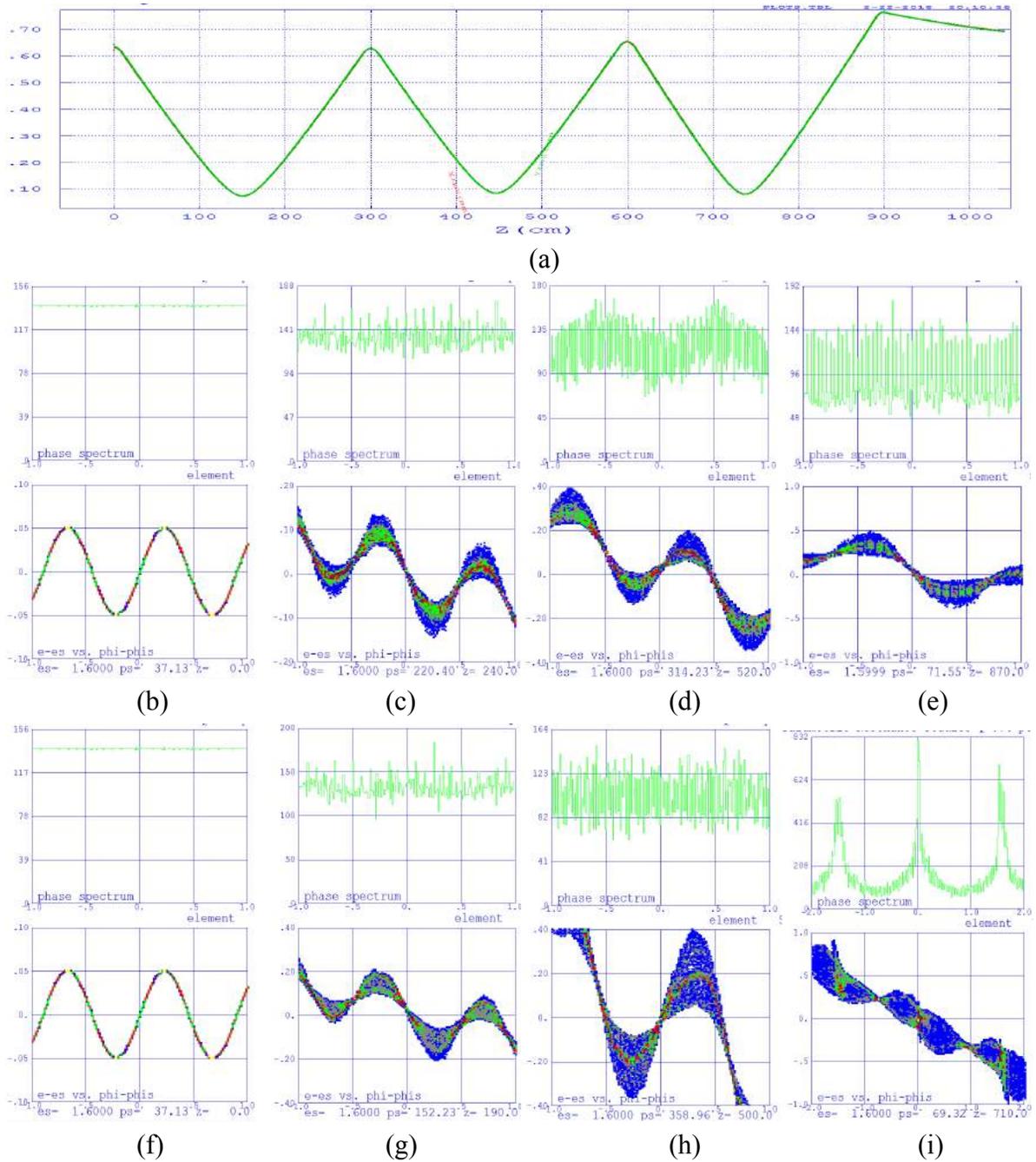

Fig. 15. PARMELA simulation of PCA instability at 4 THz for electron beam parameters in Table 3: (a) – beam envelope in mm evolution in 3-cell system; (b)-(e) Longitudinal density (top) and longitudinal phase-space (lower) plots for peak current 50 mA at z=0, 2.4, 5.2, and 8.7 meters, correspondingly, (f)-(i) Longitudinal density (top) and longitudinal phase-space (lower) plots for peak current 100 mA at z=0, 1.9, 5.0, and 7.1 meters, correspondingly.

Table 3. Parameters for PARMELA simulations

| Parameter | 50 mA | 100 mA |
|---|---|---|
| Peak current, A | 0.05 | 0.1 |
| $\gamma_o$ | 4.1 | 4.1 |
| Normalized transverse emittance, μm | 0.45 | 0.45 |
| Relative energy spread, RMS | $1\cdot10^{-3}$ | $1\cdot10^{-3}$ |
| Distance between solenoid, m | 3 | 3 |
| Transverse distribution | K-V | K-V |
| Beam radius in the waist, mm | 0.15 | 0.15 |
| $k_{sc}$ | 3 | 4.3 |
| $k_\beta$ | 7.5 | 7.5 |

## VI. Discussions.

The PCA microbunching instability we describing in this paper can be both useful and harmful. This instability does not require bending of the electron beam trajectory[7] can occur in ballistic compression. Hence it can destructive for generating high quality electron beams and, therefore, should considered as possible culprit during design process.

On positive side, PCA can be an easy way of generating nearly 100% density modulation at THz frequencies in low energy beams, which than can efficiently generate broadband radiation at these frequencies. Our analytical studies and simulations clearly indicted that such process would require electron beams with very modest parameter. At modest energies ~100 MeV and above, PCA can be used for generating intense broadband IR radiation. In short, this instability can lead to creation of new compact, inexpensive sources of high power broad-band THz and IR sources.

Our interest in this instability is driven by using it as broadband amplifier of interaction between hadrons and electrons in Coherent electron Cooling systems. The nature of plasma oscillations is that they dramatically slow down with increase of the beam energy. It is clearly indicated by scaling of dimensionless space charge parameter

$$k_{sc} = \sqrt{\frac{2}{\beta_o^3 \gamma_o^3} \frac{I_o}{I_A} \frac{l^2}{a_o^2}} \sim \gamma_o^{-1.5}.$$

---

[7] Microbunching instability caused by space charge (or coherent synchrotron radiation) and by increased longitudinal mobility (called $R_{56}$ in accelerator lingo) originating form bending of electron beam trajectory in bunch compressors (such as chicaned or doglegs) is well known and well studied both theoretically and experimentally [53-68], with excellent review in [69]. The basic process for such instability consists of two main components: energy modulation originated from space charge and longitudinal particle's slippage originated from $R_{56}$. This process does not require density modulation of the electron beam – this role is played by bending magnets which modulate longitudinal mobility of particles.

While higher energy hadron accelerators have much longer straight sections available for PCA and peak current can be increased with the beam energy, it is hard to imaging a high gain PCA working for proton beams with energy above 1 TeV. Still, it does not preclude using PCA-based CeC as a cost effective cooler for a future EIC where maximum energy of proton beam is limited to 275 GeV. We made estimations for cooling time for a typical 275 GeV proton beam in eRHIC), the BNL's options for EIC, having bunch length ~5 cm, relative energy spread ~ $2 \cdot 10^{-4}$ and RMS normalized emittance ~ 0.5 mm mrad. The PCA-based CeC driven by electron beam with 5 nC per bunch, and other parameters specified in third column of Table 1, would provide the full-beam cooling time ~ 10 minutes (with local cooling time ~ 1 minute). Cooling time for ions should be scaled down by $Z^2/A$ and will be well under one minute. With this cooling time, the PCA bandwidth and RMS bunch length of 5 cm, the number of hadrons per bunch is limited to $3 \cdot 10^{12}$, which is by order of magnitude above design values for eRHIC.

If necessary, the cooling rate can be further improved by increasing the PCA gain. The fundamental limitation for cooling time for a typical eRHIC $10^{11}$ protons per bunch and 5cm RMS bunch length will be ~ 20 seconds (local time ~ 2 second). This limit is imposed by the PCA bandwidth.

Further improvements, if needed, will require increasing the electron bunch length above the $1/10^{th}$ of the hadron bunch length used for the above estimation. It either should come from increasing charge per bunch, which can be challenge of accelerator-driver, or from deducing the peak current of electron beam. The later will cause reduction of the PCA by reduced vaue of space charge parameter, $k_{sc}$. The gain can be recuperate using more advanced PCA system shown in Fig. 4. As shown in Fig. 16, using even a small portion of the cell (from 1/5h to $1/3^{rd}$) as "strong focusing" element can significantly increase PCA gain at low space charge parameters.

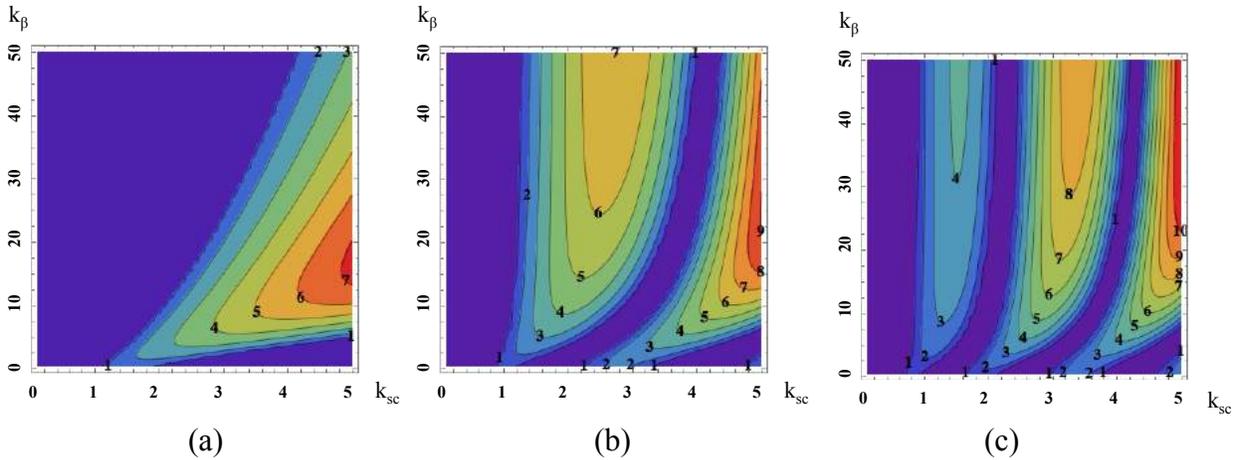

Fig. 16. Comparison of PCA cell eigen values for a regular scheme (a) and scheme with dedicated central section (Fig. 4(a)) maintaining constant beam size: fig. (b) shows the case when the central section occupies $1/5^{th}$ of the cell and fig. (c) for the case of $1/3^{rd}$. We used similar program as shown in Appendix A, which included additional "focusing element".

As indicated by our analytical simulations, this more advanced scheme allows for attaining significantly higher PCA gain at much lower beam currents. For example, with filling factor of $1/5^{th}$ (Fig. 16(b)) the eigen value of $\lambda=-6$ per cell can be reached at $k_{cs}=2$, while standard cell it would require $k_{cs}>4$, e.g. at least 4 times higher peak current. Using $1/3^{rd}$ of the cell for the additional focusing allows exponential gain at $k_{cs}\sim1$, which can reach $\lambda=-4$ per cell at $k_{cs}\sim1.4$. It also brings additional amplification bends, for example at $k_{cs}=3.2$ have positive eigen value $\lambda=8$.

In short, more advanced schemes provide for additional flexibility and potential for improvement, but expending range of parameters also require more time to finding optimal solutions for PCA-based CeC cooler. Accurate estimation of its performance would also require a dedicated set of numerical simulations, similar to that done for FEL-based CeC.

## VII. Conclusions and Acknowledgements.

In this paper we describe a novel instability occurring in linear transport beamlines, which nick-named Plasma Cascade Amplifier (PCA) of microbunching. As any other beam instabilities CPA instability can be either harmful or useful.

We propose to use CPA as a broad-band amplifier for Coherent electron Cooling of hadron beams. Our estimation show that it can cool hadron beams with sub-TeV energies on a very reasonable time scale of few minutes. This scheme naturally scales to lower energies and can be used for hadron beams with energy of few tens of GeV. We propose to use existing CeC system at RHIC (BNL) to test this novel cooling technique.

On the other hand, the CPA instability definitely can be used for generation of high power broadband THz or IR radiation or similar applications. But it also can be a dangerous instability developing in injectors for high quality electron beam required for advanced applications, such as X-Ray FELs or drivers for plasma-wakefield accelerators.

Authors would like to acknowledge genuine interest to this phenomena and continuous support by Dr. Thomas Roser (BNL). First author also would like to thank Prof. Pietro Musumeci (UCLA), who mentioned during our discussion about energy conservation in longitudinal plasma oscillations, that modulation of the transverse beam size can violate this perception. His notion was one of the initial signal that modulation of the transverse beam size can cause an instability.

This research was supported by and NSF grant PHY-1415252, , DOE NP office grant DE-FOA-0000632, and by Brookhaven Science Associates, LLC under Contract No. DEAC0298CH10886 with the U.S. Department of Energy.

# REFERENCES

[1] *"Challenges and Goals for Accelerators in the XXIst Century"*, Edited by S. Myers and O. Bruning Published by World Scientific Publishing Co. Pte. Ltd., 2016. ISBN

[2] V.N. Litvinenko, Potential for polarized luminosity increases at RHIC with CeC, BNL, July 31, 2009

[3] *"Electron-Ion Collider: The next QCD frontier"*, A. Accardi, J. L. Albacete, M. Anselmino, N. Armesto, E. C. Aschenauer et al., The European Physical Journal A, September 2016, **52**, 268, http://link.springer.com/article/10.1140/epja/i2016-16268-9

[4] LHC design report, http://documents.cern.ch/cgi-bin/

[5] "*A Large Hadron Electron Collider at CERN"*, J. L. Abelleira Fernandez t, C. Adolphsen, A.N. Akay, H. Aksakal, J.L. Albacete et al., Journal of Physics G: Nuclear and Particle Physics, Volume 39, Number 7, July 2012, pp. 75001-75630 (630) http://iopscience.iop.org/0954-3899/39/7/075001

[6] A. Caldwell, K. Lotov, A. Pukhov, and F. Simon, Nat. Phys. 5, 363 (2009).

[7] [3] G. I. Budker, Sov. At. Energy 22, 438 (1967)

[8] S.Nagaitsev et al., Phys. Rev. Lett. 96, 044801 (2006)

[9] S. van der Meer, Rev. Mod. Phys. 57, 689 (1985)

[10] M.Blaskiewicz, J.M. Brennan, F Severino, Physical Review Letters 100, 174802 (2008)

[11] A.Mikhalichenko, M.Zolotorev, Phys. Rev. Lett., 71, p.4146 (1993)

[12] V.N. Litvinenko, Y.S. Derbenev, *Coherent Electron Cooling*, Phys. Rev. Lett. **102**, 114801 (2009)

[13] D. Ratner, *Microbunched Electron Cooling for High-Energy Hadron Beams*, Phys . Rev. Lett. **111** 084802 (2013)

[14] V.N. Litvinenko, Y.S. Derbenev, Proc. of 29th International Free Electron Laser Conference, Novosibirsk, Russia, 2007, p. 268. http://accelconf.web.cern.ch/accelconf/f07/PAPERS/TUCAU01.PDF

[15] *Advances in Coherent Electron Cooling*, V.N. Litvinenko, In Proceedings of COOL 2013 workshop, June 2013, Mürren, Switzerland, p. 175, ISBN 978-3-95450-140-3, http://accelconf.web.cern.ch/AccelConf/COOL2013/papers/tham2ha02.pdf

[16] G. Stupakov, *"Derivation of the cooling rate for Microbunched Electron Cooling (MBEC)"*, SLAC-PUB-17208 (2018).

[17] Ya. S. Derbenev, in Proceedings of 7th Conference on Charged Particle Accelerators, Dubna, USSR, 1980, p. 269

[18] Ya. S. Derbenev, in Proceedings of AIP **253** (AIP, College Park, MD, 1992), pp. 103–110.

[19] Ya. S. Derbenev, "Coherent Electron Cooling", UM HE 91-28 (1991) http://inspirehep.net/record/318036
ibid.  https://lib-extopc.kek.jp/preprints/PDF/2000/0035/0035838.pdf

[20] Y. Hao, V. Litvinenko, S*imulation Study of Electron Response Amplification in Coherent Electron Cooling*, In Proc. of Third International Particle Accelerator Conference, New Orleans, USA, May 20 - 25, 2012, p. 448

[21] E.L. Saldin, E.A. Schneidermiller, M.V. Yurkov, The Physics of FELs, Springer, 1999

[22] V.N. Litvinenko et al., "Proof-of-principle Experiment for FEL-based Coherent Electron Cooling", THOBN3, PAC'11, New York, NY, 2011

[23] V.N. Litvinenko *et al.*, "Commissioning of FEL-based Coherent electron Cooling system", In *Proc. of FEL 2017,* Santa Fe*,* NM, 2017

[24] I. Pinayev *et al.*, "Commissioning of CeC PoP Accelerator", in *Proc. North American Particle Accelerator Conf. (NAPAC'16)*, Chicago, IL, USA, Oct. 2016, paper WEPOB60, pp. 1027-1029, https://jacow.org/napac2016/papers/wepob60.pdf , 2017.

[25] D.R. Nicholson, Introduction in Plasma Theory, John Wiley & Sons, 1983

[26] see, for example, Advanced Accelerator Physics, http://case.physics.stonybrook.edu/index.php/PHY564_fall_2017

[27] http://case.physics.stonybrook.edu/images/a/a3/PHY564_Lecture_4_F2017.pdf, p.35-38

[28] M. Reiser, Theory and design of charged particle beams, John Wiley & Sons, 2008.

[29] E.D. Courant, H.S. Snyder, Ann. Phys. 3 (1958) 1.

[30] E. Forest, "Forth-order symplectic integrator", SLAG-PUB-5071, LBL-27662, August 1989

[31] Mathematica, © Wolfram Research, Inc

[32] M. Abramowitz and I. A. Stegan, Handbook on mathematical Functions, National Bureau of Standards, 1964, p. 555

[33] X. Wang, R. Samulyak, J. Jiao, K. Yu, Adaptive Particle-in-Cloud method for optimal solutions to Vlasov-Poisson equation, J. Comput. Phys., 316 (2016), 682 - 699.

[34] L.M. Young, J.H. Billen, PARMELA documentation, LA-UR-96-1835.

[35] J. Ma, Numerical Algorithms for Vlasov-Poisson Equation and Applications to Coherent Electron Cooling, Ph.D. Thesis, State University of New York at Stony Brook, ProQuest Dissertations Publishing, 2017.

[36] J. Ma, G. Wang, R. Samulyak, V.N. Litvinenko, X. Wang, K. Yu, SIMULATION STUDIES OF MODULATOR FOR COHERENT ELECTRON COOLING, submitted to Physical Review Accelerators and Beams

[37] IMPACT-T: A 3D Parallel Particle Tracking Code in Time Domain, Ji Qiang, LBNL, http://amac.lbl.gov/~jiqiang/IMPACT-T/index.html

[38] Klaus Floettmann, DESY, ASTRA - A Space Charge Tracking Algorithm, http://www.desy.de/~mpyflo/Astra_manual/Astra-Manual_V3.2.pdf

[39] G.Wang, M.Blaskiewicz, Phys Rev E, volume 78, 026413 (2008)

[40] *Dynamics of shielding of a moving charged particle in a confined electron plasma,* Andrey Elizarov and Vladimir Litvinenko, Phys. Rev. ST Accel. Beams 18, 044001*,* 2015*,* https://journals.aps.org/prstab/pdf/10.1103/PhysRevSTAB.18.044001

[41] Analytical Studies of Ion Beam Evolution under Coherent Electron Cooling, G. Wang, M. Blaskiewicz, V.N. Litvinenko, In Proceedings of IPAC 2016, May 8-13, 2016, Busan, Korea, 2016, p. 1247, http://accelconf.web.cern.ch/AccelConf/ipac2016/papers/tupmr009.pdf

[42] Modulator Simulations for Coherent Electron Cooling, Jun Ma, Gang Wang, Xingyu Wang, Roman Samulyak,, Vladimir N. Litvinenko, Kwangmin Yu, Proceedings of NA-PAC, October 9 - 14, 2016, Chicago, IL, USA

[43] Simulation of Ion Beam under Coherent Electron Cooling, G. Wang, M. Blaskiewicz, V. Litvinenko, In Proceedings of IPAC 2016, May 8-13, 2016, Busan, Korea, TUPMR008, p. 1243-1246

[44] G. Wang, Coherent Electron Cooling and Two Stream Instabilities Due to Electron Cooling, PhD, Stony Brook University, December 2008

[45] G. Stupakov and M. S. Zolotorev, Phys. Rev. Lett. **110**, 269503 (Jun 2013)

[46] V.N. Litvinenko and Ya. S. Derbenev, Phys. Rev. Lett. 110, 269504, (June 2013)

[47] G. Stupakov and M. S. Zolotorev, *Reevaluation of Coherent Electron Cooling Gain Factor*, in Proceedings of the 2013 FEL Conference, New York, USA, 2013.

[48] *Model-independent Description of Shot-noise, Amplification and Saturation*, Y.C. Jing, V. Litvinenko, G. Wang, Proceedings of Fifth International Particle Accelerator Conference, Dresden, Germany, IPAC 2014, June 15-20, 2014, p. 2949, http://accelconf.web.cern.ch/AccelConf/IPAC2014/papers/thpro039.pdf

[49] Y. Jing, V.N. Litvinenko, Y. Hao, G. Wang, Model Independent Description of Amplification and Saturation Using Green's Function, arXiv preprint arXiv:1505.04735, (2015)

[50] *Analysis of FEL-based CeC Amplification at High Gain Limit*, Gang Wang, Yichao Jing, Vladimir N. Litvinenko, THPF144.PDF, In Proc. of IPAC 2015, May 3-8, 2015, Richmond, VA

[51] A.V. Fedotov, M. Blaskiewicz, W. Fischer, D. Kayran, J. Kewisch et al. " Accelerator Physics Design Requirements and Challenges of RF Based Electron Cooler LEReC", In Proc. of NAPAC2016, Chicago, IL, USA, 2016, p. 867

[52] R.A. Lacey, Phys. Rev. Lett. **114**, 142301, 2015

[53] E. L. Saldin, E. A. Schneidmiller, and M.V. Yurkov, Nucl. Instrum. Methods Phys. Res., Sect. A 490, 1 (2002).

[54] S. Heifets, G. Stupakov, and S. Krinsky, Phys. Rev. ST Accel. Beams 5, 064401 (2002).

[55] E. A. Schneidmiller and M.V. Yurkov, Phys. Rev. ST Accel. Beams 13, 110701 (2010)

[56] T. Shaftan and Z. Huang. Phys. Rev. ST-AB, 7:080702, 2004.

[57] Marco Venturini, Phys. Rev. ST Accel. Beams 10, 104401 (2007).

[58] M. Venturini, *Models of longitudinal space-charge impedance for microbunching instability*. Phys. Rev. ST Accel. Beams, 11:034401, 2008.

[59] D. Ratner, Z. Huang, A. Chao, *Three-Dimensional Analysis of Longitudinal Space Charge Microbunching Starting From Shot Noise* Presented at FEL08, Gyeongju, Korea, 24-29 Aug (2008)

[60] R. Akre et al., Phys. Rev. ST Accel. Beams 11, 030703 (2008)

[61] A. Marinelli and J.B. Rosenzweig, *Microscopic kinetic analysis of space-charge induced optical microbunching in a relativistic electron beam*, Phys. Rev. ST Accel. Beams 13, 110703 (2010)

[62] A. Marinelli, E. Hemsing, M. Dunning, D. Xiang, S. Weathersby, F. O'Shea, I. Gadjev, C. Hast, and J. B. Rosenzweig, Phys. Rev. Lett. 110, 264802 (2013).

[63] Zhirong Huang and Kwang-Je Kim, Phys. Rev. ST Accel. Beams 5, 074401 (2002).

[64] Z. Huang, M. Borland, P. Emma, J. Wu, C. Limborg, G. Stupakov, and J. Welch, Phys. Rev. ST Accel. Beams 7, 074401 (2004).

[65] D. Ratner. Much Ado About Microbunching: Coherent Bunching in High Brightness Electron Beams, PhD thesis, Department of Applied Physics, Stanford University, 2011.

[66] A. Marinelli, E. Hemsing, and J.B. Rosenzweig, *Three dimensional analysis of longitudinal plasma oscillations in a thermal relativistic electron beam*, Physics of Plasmas, 18:103105, 2011.

[67] V.N. Litvinenko, G. Wang, *Relativistic Effects in Micro-Bunching*, Proceedings of FEL2014, Basel, Switzerland, August 25-29, 2014, THP035, http://www.fel2014.ch/prepress/FEL2014/papers/thp035.pdf

[68] Gain limitation in micro-bunching amplifier, Notes written by V.N. Litvinenko and G. Stupakov, December 2017

[69] D. Ratner, *Microbunching Instability: Review of Current Models, Observations, and Solutions,* ICFA Beam Dynamics Newsletter No. 49 p. 112 (2012)

Appendix A. **Mathematica program**

We used Mathematica [31] for evolution of the eigen values of the PCA cell. Typical structure is shown bellow. We are performing calculations for *k1step* steps in $k_{sc}$ and *k2step* steps in $k_\beta$. At each point we split half of cell $\hat{s} \in \{0,1\}$ interval into *nstep* equal intervals and set initial condition of

$$\hat{a}(0)=1;\ p=\hat{a}'(0)=0;\ \mathbf{M}(0|0)=\begin{bmatrix} 1 & 0 \\ 0 & 1 \end{bmatrix}. \quad (A.1)$$

At each step in $\hat{s}$ we used calculate the envelope evolution using 4-th order symplectic integrator. After that we calculate transport matrix of this interval and multiply it from left with $\mathbf{M}$. When process is finished at $\hat{s}$ we obtain Matrix of a half-cell $m = \mathbf{M}(0|1)$. Using bilateral symmetry of the cell we then calculating the transport matrix of the cell $t = \mathbf{M}(-1|1)$:

$$\begin{aligned}
\mathbf{M}(-1|1) &= \mathbf{M}(0|1)\mathbf{M}(-1|0); \\
\mathbf{M}(-1|0) &= \begin{bmatrix} a & b \\ c & d \end{bmatrix}; ad-bc=1; \\
\mathbf{M}(-1|0) &= \tilde{\mathbf{M}}(0|1) = \begin{bmatrix} d & b \\ c & a \end{bmatrix}; \\
\mathbf{M}(-1|1) &= \begin{bmatrix} ad+bc & 2ab \\ 2dc & ad+bc \end{bmatrix} = \begin{bmatrix} m_{11} & m_{12} \\ m_{21} & m_{11} \end{bmatrix}.
\end{aligned} \quad (A.2)$$

```
nstep = 200; ds = 1.0/nstep;
α = 1. - 2.^(1/3); A = 1/(1+α); Ads = A*ds;
k1step = 100; k2step = 100; kscmax = 20; kβmax = 20;
λ = Table[0, {i, 1, k2step}, {j, 1, k1step}];
λ2t = Table[0, {i, 1, k2step}, {j, 1, k1step}];
a = Table[0.0, {i, 1, nstep+1}]; p = Table[0.0, {i, 1, nstep+1}];
Do[ksc = kscmax*k/k1step; c1 = ksc^2; Do[κ = kβmax*j/k2step; c2 = κ^2;
  a[[1]] = 1.; p[[1]] = 0; m = {{1., 0.}, {0., 1.}}; φp = 0.;
  Do[aa = a[[i]]; pp = p[[i]];
    pp = pp + (c1/aa + c2/aa^3)*Ads/2;
    aa = aa + pp*Ads;
    pp = pp + (c1/aa + c2/aa^3)*α*Ads/2;
    aa = aa + pp*(α-1)*Ads;
    pp = pp + (c1/aa + c2/aa^3)*α*Ads/2;
    aa = aa + pp*Ads;
    pp = pp + (c1/aa + c2/aa^3)*Ads/2;
    p[[i+1]] = pp; a[[i+1]] = aa;
    aa = 0.5*(a[[i]] + a[[i+1]]);
    ω = Sqrt[2*c1/aa^2]; φ = ω*ds; φp = φp + φ;
    m1 = {{Cos[φ], Sin[φ]/ω}, {-Sin[φ]*ω, Cos[φ]}};
    m = m1.m, {i, 1, nstep}];
  mt = {{m[[2, 2]], m[[1, 2]]}, {m[[2, 1]], m[[1, 1]]}};
  t = m.mt;
  λ1 = t[[1, 1]] - Sqrt[t[[1, 1]]^2 - 1.];
  λ2 = t[[1, 1]] + Sqrt[t[[1, 1]]^2 - 1.]; λ[[j, k]] = Max[Abs[Re[λ1]], Abs[Re[λ2]]],
  {j, 1, k2step}], {k, 1, k1step}]
```

Finally we calculate eigen values of the cell matrix

$$\lambda_{1,2} = m_{11} \mp \sqrt{m_{11}^2 - 1} \qquad (A.3)$$

and evaluate maximum of the absolute values of their real parts:

$$\lambda = \max\left(\left|\operatorname{Re}\lambda_1\right|, \left|\operatorname{Re}\lambda_2\right|\right) \qquad (A.4)$$

If $\lambda < 1$, plasma oscillations are stable, and if $\lambda > 1$, the amplitude of the plasma oscillations growing exponentially as described in eqs. (6-8).

For simulations of the central section with constant beam raduaus (scheme shown in Fig. 4(b), we had change initial condition for the initial matrix

$$m_{s=0} = \begin{bmatrix} \cos\varphi_o & \dfrac{\sin\varphi_o}{\omega_o} \\ -\omega_o \sin\varphi_o & \cos\varphi_o \end{bmatrix}; \omega_o = \sqrt{2}k_{sc}; \varphi_o = \eta\omega_o; \eta = \dfrac{l_s}{2l}; \qquad (A.5)$$

and followed the rest of the process as in program above.

Appendix B. **Cooling rate estimations**

  B.1 *Output from the CeC modulator*

  First, we derive the input into the density modulation using technique described in [39,44]. This section describes many of the processed in the beam frame and many notation introduced in [39,44], which are different from those used in the main text of the paper. According to [44] the density modulation in frequency domain reads

$$\tilde{n}_1(\vec{k},t) = \frac{Z_i \omega_p^2}{\omega_p^2 + \lambda(\vec{k})^2}\left[1 - e^{\lambda(\vec{k})t}\left(\cos(\omega_p t) - \frac{\lambda(\vec{k})}{\omega_p}\sin(\omega_p t)\right)\right], \qquad (B.1)$$

with

$$\lambda(\vec{k}) = i\vec{k}\cdot\vec{v}_0 - \sqrt{(k_x \beta_x)^2 + (k_y \beta_y)^2 + (k_z \beta_z)^2} \quad . \qquad (B.2)$$

The inverse Fourier transformation of eq. (B.1) over the transverse planes is

$$\tilde{n}_1(k_z,x,y,t) = \frac{1}{(2\pi)^2}\int_{-\infty}^{\infty} e^{ik_x x} e^{ik_y y} \tilde{n}_1(\vec{k},t)\, dk_x\, dk_y \,, \qquad (B.3)$$

and integrating eq. (B.3) over the transverse planes gives the Fourier transformation of the line density perturbation:

$$\begin{aligned}\tilde{\rho}(k_z,t) &= \frac{1}{(2\pi)^2}\int_{-\infty}^{\infty} e^{ik_x x} e^{ik_y y} \tilde{n}_1(\vec{k},t)\, dk_x\, dk_y\, dx\, dy = \int_{-\infty}^{\infty}\delta(k_x)\delta(k_y)\tilde{n}_1(\vec{k},t)\, dk_x\, dk_y \\ &= \frac{Z_i\omega_p^2}{\omega_p^2 + \lambda_z(k_z)^2}\left[1 - e^{\lambda_z(k_z)t}\left(\cos(\omega_p t) - \frac{\lambda_z(k_z)}{\omega_p}\sin(\omega_p t)\right)\right]\end{aligned} \qquad (B.4)$$

with

$$\lambda_z(k_z) = ik_z v_{0z} - |k_z|\beta_z .$$ (B.5)

Hence, at the exit of the modulator, the Fourier components of the line charge density perturbation is

$$\tilde{\rho}_1(k_z) = \frac{Z_i e}{1+\bar{\lambda}_z(k_z)^2}\left[1 - e^{\bar{\lambda}_z(k_z)\psi_m}\left(\cos(\psi_m) - \bar{\lambda}_z(k_z)\sin(\psi_m)\right)\right],$$ (B.6)

with

$$\bar{\lambda}_z(k_z) = \frac{\lambda_z(k_z)}{\omega_p},$$ (B.7)

and $\psi_m$ is the phase advance of plasma oscillation in the modulator

$$\psi_m = 2\pi \frac{L_m}{\lambda_{p,lab}}.$$ (B.8)

The inverse Fourier transformation of eq. **Error! Reference source not found.** can be carried out, which gives the following expression:

$$\rho_{1z}(z) = \frac{Z_i e}{\pi a_z}\int_0^{\psi_m}\frac{t_1 \sin t_1}{\left(\frac{z}{a_z}+\frac{v_{0z}}{\beta_z}t_1\right)^2 + t_1^2},$$ (B.9)

where

$$a_z = \frac{\beta_z}{\omega_p}.$$ (B.10)

is longitudinal Debye length in beam frame.

B.2. *Output from plasma cascade amplifier*
For the initial modulation of the form

$$\tilde{f}_1(k_z,v_z,0) = \frac{1}{\pi\beta_z}\frac{\tilde{\rho}_1(k_z)}{1+\frac{v_z^2}{\beta_z^2}},$$ (B.11)

the output of the line density perturbation at the exit of the amplifier reads

$$\tilde{\rho}_2(k_z) = g_{amp}\exp\left(-|k_z|\beta_z\frac{L_{amp}}{\gamma c}\right)R_{amp}(k_z)\tilde{\rho}_1(k_z),$$ (B.12)

where $g_{amp}$ is the amplification gain (independent of $k_z$), $L_{amp}$ is the length of the amplifier and (see APPENDIX C)

$$R_{amp}(k_z) = k_z^2 a_{amp}^2 \int_0^1 \eta K_0(|k_z| a \cdot \eta) d\eta, \tag{B.13}$$

is a reduction factor due to finite beam width. To simplify the evaluation, we can also use an approximate form for the reduction factor as

$$R_{app}(k_z) = \frac{k_z^2 a_{amp}^2}{1.4 + k_z^2 a_{amp}^2}. \tag{B.14}$$

Fig. B1. shows the comparison of the reduction factor from the exact integral, eq. (B.13), and that from the approximate expression from eq. (B.14). The approximate form, eq. (B.14), is useful when one tries to get close form solution for the inverse Fourier transformation, but since we are doing numerical evaluation anyway, we will use eq. (B.13) to evaluate the amplified density perturbation, i.e.

$$\tilde{\rho}_2(k_z) = g_{amp} \exp\left(-|k_z|\beta_z \frac{L_{amp}}{\gamma c}\right) R_{amp}(k_z) \tilde{\rho}_1(k_z). \tag{B.15}$$

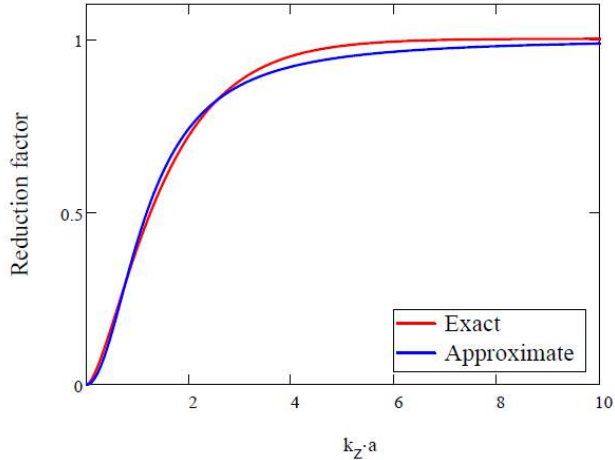

Figure B.1: Comparison of reduction factor from integration of modified Bessel function (eq. (B.13)) and from the approximate expression of eq. (B.14).

As seen from Fig. B.1, the Fourier component is zero at $k_z = 0$. This property guarantees the charge conservation since

$$\tilde{\rho}_1(k_z)\big|_{k_z=0} = \int_{-\infty}^{\infty} \rho_1(z) e^{-ik_z z} dz \bigg|_{k_z=0} = \int_{-\infty}^{\infty} \rho(z) dz = \Delta Q = 0. \tag{B.16}$$

The observation of eq. (B.16) is also confirmed by the appearance of negative values for the blue curve in Fig. 3.B.

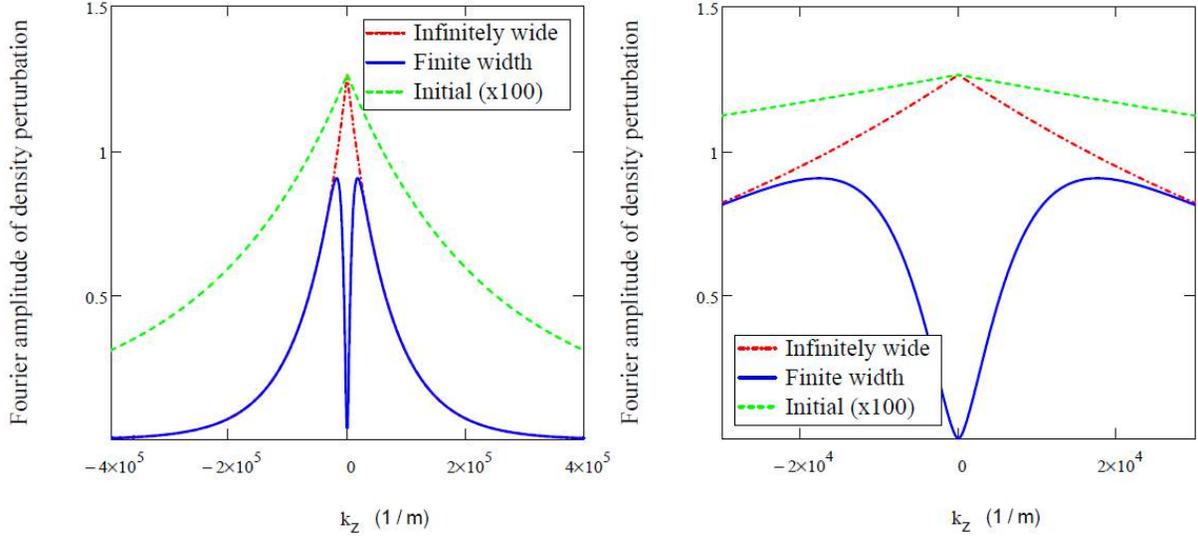

Figure 2.B: Fourier amplitude of charge density perturbation before (green) and after amplification (red and blue). The ordinate is the line charge density perturbation in unit of $10^{-15}C$. The abscissa is the value of the longitudinal wave-vector, $k_z$, in the beam frame. The green curve is the initial perturbation at the exit of modulator as calculated from eq. (B.6). The red curve is the charge density perturbation after amplification for an infinitely wide beam, i.e. $R_{amp} = 1$. The blue curve is the charge density perturbation after amplification for a beam with finite size ( $a_{amp} = 0.2mm$ is used in the plot). The peak of the blue curve locates at $k_z = 17580/m$, which corresponds to lab frame frequency of $f_{peak} = k_{peak} c /(2\pi) = 2.4 \times 10^{13} Hz$.

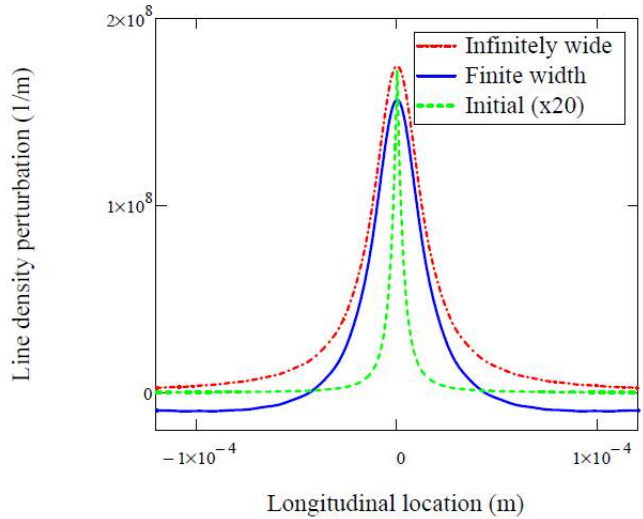

Figure 3.B: Line number density perturbation before and after amplification (red and blue). The abscissa is the longitudinal location along the electron beam in beam frame. The green curve is the initial density perturbation amplified by a factor of 20 for visibility, the red curve is the density perturbation after amplification for an infinitely wide electron beam and the blue is the line density perturbation after amplification for a beam with finite width ( $a_{amp} = 0.2mm$ ).

B.3. *Longitudinal electric field in CeC kicker*

The longitudinal electric field as a function of radial offset, for $r < a$, is given by (see APPENDIX C)

$$\tilde{E}_z(k_z,r) = -\frac{\tilde{\rho}_2(k_z)}{\pi\varepsilon_0} ik_z \left[ I_0(k_z r) \int_{r/a}^{1} \eta K_0(k_z a \cdot \eta) d\eta + K_0(k_z r) \int_0^{r/a} \eta I_0(k_z a \cdot \eta) d\eta \right], \quad (B.17)$$

where $a$ is the radius of the electron beam at the kicker section. The longitudinal electric field in the time domain is thus given by

$$E_z(z,r) = \frac{1}{2\pi} \int_{-\infty}^{\infty} \tilde{E}_z(k_z,r) e^{ik_z z} dk_z = \frac{1}{\pi} \int_0^{\infty} \mathrm{Re}\left[ \tilde{E}_z(k_z,r) e^{ik_z z} \right] dk_z. \quad (B.18)$$

The second equation of eq. (B.18) results from the fact that

$$\tilde{E}_z^*(k_z,r) = \int_{-\infty}^{\infty} E_z(z,r) e^{ik_z z} dz = \tilde{E}_z(-k_z,r) \quad (B.19)$$

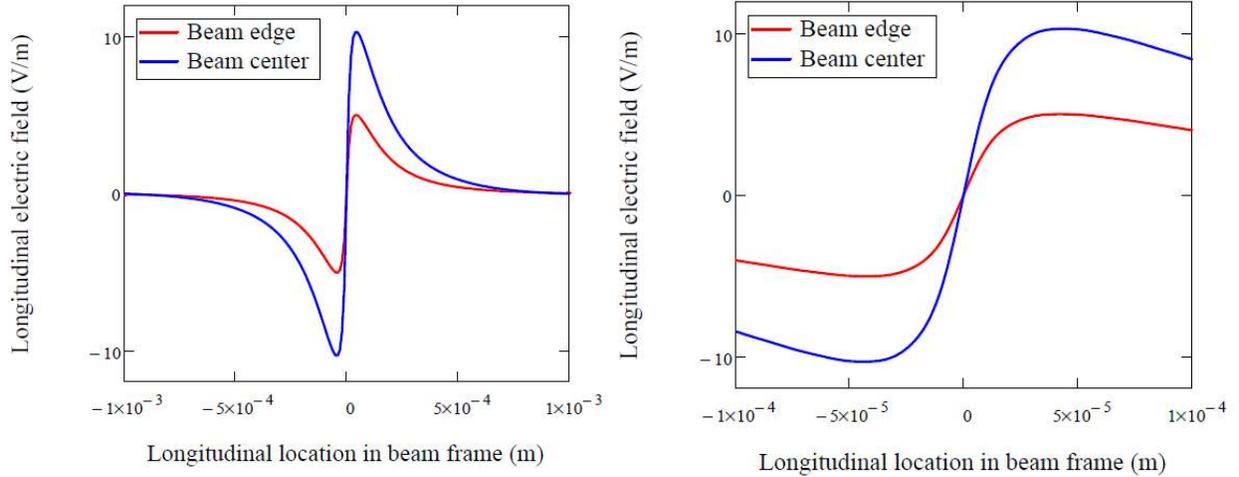

Figure 2: Longitudinal electric field at the kicker as a function of longitudinal distance from the initial modulation. The ordinate is the longitudinal electric field at beam transverse center (blue) and beam edge, $r = a$ (red), as calculated from eq.(B.18). The parameters used in this plot are $Z_i = 79$, $\psi_m = \pi/2$, $L_{amp} = 8m$, $I_{peak} = 100A$, $a = 1.1mm$, $\gamma_e = 28.55$ and $\delta_p = 10^{-4}$.

B.4. *Estimates of cooling time*

As an estimate, where we take an average field between the beam center and its edge of

$$E_{peak} = 10 V/m, \quad (B.20)$$

and with plasma wavelength of about 7 meters we assume the kicker length of 3 meters:

$$L_k = 3m \quad (B.21)$$

the peak energy kick per pass is

$$\Delta E_{correction} = Z_i e E_{peak} L_k = 2.37 KeV. \quad (B.22)$$

Assuming the RMS energy spread of the ion is

$$\sigma_{\delta,rms} = 6 \times 10^{-4} , \tag{B.23}$$

the local cooling time can be estimated as

$$T_{cool} = \frac{T_{rev}\sigma_{\delta,ion}E_{ion}}{\Delta E_{correction}} \approx 17s , \tag{B.24}$$

with the energy of the ion given by

$$E_{ion} = A_i m_u \gamma_{ion} c^2 \tag{B.25}$$

and $A_i = 197$ for Au ion.

Appendix C. **Periodic longitudinal electric field in finite electron beams.**
The Poisson equation for charge distribution of the form

$$\rho(r,\theta,z) = \rho_0 f_{//}(z) f_\perp(r) \tag{C.1}$$

reads

$$\frac{1}{r}\left[\frac{\partial}{\partial r}\left(r\frac{\partial}{\partial r}\varphi(r,z)\right)\right] + \frac{\partial^2}{\partial z^2}\varphi(r,z) = \frac{\rho_0}{\varepsilon_0} f_{//}(z) f_\perp(r) , \tag{C.2}$$

where $f_\perp(r)$ is the transverse surface density of electrons in unit of $m^{-2}$, $\rho_0$ is perturbation of line charge density in unit of $C/m$ and $f_{//}(z)$ is the normalized perturbative line charge distribution. Multiplying both sides of eq. (C.2) by $e^{-ikz}$ and integrating over $z$ yields

$$\frac{1}{r}\left[\frac{\partial}{\partial r}\left(r\frac{\partial}{\partial r}\phi\right)\right] - k^2\phi = f(r) , \tag{C.3}$$

which can be written as

$$\frac{\partial^2}{\partial r^2}\phi + \frac{1}{r}\frac{\partial}{\partial r}\phi - k^2\phi = f(r), \tag{C.4}$$

with

$$\phi(k,r) = \int_{-\infty}^{\infty} e^{-ikz}\varphi(r,z)dk , \tag{C.5}$$

$$\varphi(r,z) = \frac{1}{2\pi}\int_{-\infty}^{\infty} \phi(k,r)e^{ikz}dz , \tag{C.6}$$

$$f(r) = \frac{\rho_0}{\varepsilon_0} f_\perp(r)\int_{-\infty}^{\infty} e^{-ikz} f_{//}(z)dz = \frac{\rho_0}{\varepsilon_0} f_\perp(r)\tilde{f}_{//}(k) , \tag{C.7}$$

and

$$\tilde{f}_{//}(k) = \int_{-\infty}^{\infty} e^{-ikz} f_{//}(z)dz . \tag{C.8}$$

The homogenous part of eq. (C.4) is

$$\frac{\partial^2}{\partial r^2}\phi_h + \frac{1}{r}\frac{\partial}{\partial r}\phi_h - k^2\phi_h = 0, \tag{C.9}$$

which have general solution of the modified Bessel function, i.e.

$$\phi_h = c_1 I_0(kr) + c_2 K_0(kr). \tag{C.10}$$

The solution for inhomogeneous equation, eq. (C.4), can be written as[8]

$$\phi(r) = c_1 I_0(kr) + c_2 K_0(kr) + \frac{1}{k}\int_{r_0}^{r} \frac{I_0(k\xi)K_0(kr) - K_0(k\xi)I_0(kr)}{I_0(k\xi)K_0'(k\xi) - K_0(k\xi)I_0'(k\xi)} \cdot f(\xi)d\xi \tag{C.11}$$

with $c_1$ and $c_2$ being some constants determined by the initial conditions at $r = r_0$. With $r_0 = \infty$, eq. (C.11) becomes

$$\phi(r) = I_0(kr)\left[c_1 + \int_{\infty}^{r} \xi K_0(k\xi) \cdot f(\xi)d\xi\right] + K_0(kr)\left[c_2 - \int_{\infty}^{r} \xi I_0(k\xi) \cdot f(\xi)d\xi\right], \tag{C.12}$$

where we used the relation

$$I_m(x)K_m'(x) - K_m(x)I_m'(x) = -\frac{1}{x}. \tag{C.13}$$

Applying the initial condition of $\phi(\infty) = 0$ yields

$$c_1 = 0, \tag{C.14}$$

and eq. (C.12) becomes

$$\phi(r) = I_0(kr)\int_{\infty}^{r}\xi K_0(k\xi) \cdot f(\xi)d\xi + K_0(kr)\left[c_2 - \int_{\infty}^{r}\xi I_0(k\xi) \cdot f(\xi)d\xi\right]. \tag{C.15}$$

The condition $\phi'(\infty) = 0$ is automatically satisfied since

$$\lim_{r\to\infty}\frac{d}{dr}\phi(r) = c_2 k \cdot \lim_{r\to\infty}K_0'(kr) = -c_2 k \cdot \lim_{r\to\infty}K_1(kr) = 0,$$

where the relation $\frac{d}{dz}K_0(z) = -K_1(z)$ is used. The requirement of $\lim_{r\to 0}\phi(r) \neq \infty$ requires

$$\phi(0) = K_0(kr)\left[c_2 + \int_{0}^{\infty}\xi I_0(k\xi) \cdot f(\xi)d\xi\right], \tag{C.16}$$

i.e.

$$c_2 = -\int_{0}^{\infty}\xi I_0(k\xi) \cdot f(\xi)d\xi. \tag{C.17}$$

Inserting eq. (C.17) into eq. (C.15) leads to

$$\phi(r) = I_0(kr)\int_\infty^r \xi K_0(k\xi)\cdot f(\xi)d\xi - K_0(kr)\int_0^r \xi I_0(k\xi)\cdot f(\xi)d\xi. \quad \text{(C.18)}$$

---

[8] See 'Table of Integrals, Series and Products' by I. S. Gradshteyn and I.M. Ryzhik, 7$^{th}$ Edition, Chapter 16.516, pp. 1100.

Using eq. (C.5) and eq. (C.18), the longitudinal electric field is obtained as

$$E_z = -\frac{\partial \varphi}{\partial z} = -\frac{i}{2\pi}\int_{-\infty}^\infty k\phi(k,r)e^{ikz}\,dk = \frac{1}{2\pi}\int_{-\infty}^\infty \tilde{E}_z(r,k)e^{ikz}\,dk, \quad \text{(C.19)}$$

with

$$\tilde{E}_z(r,k) = -ik\phi(r) = -ik\left\{I_0(kr)\int_\infty^r \xi K_0(k\xi)\cdot f(\xi)d\xi - K_0(kr)\int_0^r \xi I_0(k\xi)\cdot f(\xi)d\xi\right\}. \quad \text{(C.20)}$$

and the transverse electric field is obtained as

$$E_r(r,z) = -\frac{\partial \varphi}{\partial r} = -\frac{1}{2\pi}\int_{-\infty}^\infty \left[\frac{\partial}{\partial r}\phi(k,r)\right]e^{ikz}\,dk = \frac{1}{2\pi}\int_{-\infty}^\infty \tilde{E}_r(r,k)e^{ikz}\,dk, \quad \text{(C.21)}$$

with

$$\tilde{E}_r = -\frac{\partial \phi}{\partial r} = kI_1(kr)\int_r^\infty \xi K_0(k\xi)\cdot f(\xi)d\xi - kK_1(kr)\int_0^r \xi I_0(k\xi)\cdot f(\xi)d\xi, \quad \text{(C.22)}$$

where we used well-known ratios

$$\frac{d}{dz}I_0(z) = I_1(z), \quad \text{(C.23)}$$

and

$$\frac{d}{dz}K_0(z) = -K_1(z). \quad \text{(C.24)}$$

If we take

$$f(r) = \frac{\rho_0}{\varepsilon_0}\tilde{f}_{//}(k)\frac{1}{\pi a^2}H(a-r), \quad \text{(C.25)}$$

with $H(x)$ being the Heaviside step function and $a$ being the beam radius and the normalization of $f(r)$ is

$$2\pi\int_0^\infty rf(r)dr = 1. \quad \text{(C.26)}$$

For $r < a$, inserting eq. (C.25) into eq. (C.20) yields

$$\tilde{E}_z(r) = -ik\frac{\rho_0}{\pi\varepsilon_0}\tilde{f}_{//}(k)\left[I_0(kr)\int_{r/a}^1 \eta K_0(ka\cdot\eta)d\eta + K_0(kr)\int_0^{r/a}\eta I_0(ka\cdot\eta)d\eta\right]. \quad \text{(C.27)}$$

To include the regions for $r > a$, the following equation is to be used

$$E_z(r) = -ik \frac{\rho_0}{\pi\varepsilon_0} \tilde{f}_{//}(k)$$

$$\times \left[ I_0(kr) \int_{r/a}^{1} \eta H(1-\eta) K_0(ka\cdot\eta) d\eta + K_0(kr) \int_{0}^{r/a} \eta H(1-\eta) I_0(ka\cdot\eta) d\eta \right]. \quad \text{(C.28)}$$

For $r = 0$, eq. (B.17) becomes

$$\tilde{E}_z(k_z) = -ik \frac{\rho_0}{\pi\varepsilon_0} \tilde{f}_{//}(k) \int_{0}^{1} \eta K_0(ka\cdot\eta) d\eta . \quad \text{(C.29)}$$